\documentclass[%
reprint,
superscriptaddress,
nobibnotes,
 amsmath,amssymb,
prb,
]{revtex4-1}

\usepackage{graphicx}
\usepackage{dcolumn}
\usepackage{bm}
\usepackage{multirow}
\usepackage{mathptmx}
\usepackage{siunitx}
\usepackage{color}
\usepackage{url}
 
\newcommand{\beq}[1]{\begin{equation}\label{#1}}
\newcommand{\eep}{\;.\end{equation}}
\newcommand{\eec}{\;,\end{equation}}
\newcommand{\eeq}{\end{equation}}

\newcommand{\s}{\sigma}

\newcommand{\D}{\Delta}
\newcommand{\eo}{{\epsilon}_0}

\newcommand{\Ep}{\mathcal{E}}

\newcommand*\dd{\mathop{}\!\mathrm{d}} 
\newcommand\del{\partial} 

\newcommand*\chem[1]{\ensuremath{\mathrm{#1}}} 

\DeclareMathAlphabet{\mathcal}{OMS}{cmsy}{m}{n} 

\begin{document}


\title{Flexoelectric radial polarization of single-walled nanotubes from first-principles}


\author{Daniel Bennett}
\email{db729@cam.ac.uk}
 \affiliation{Theory of Condensed Matter, Cavendish Laboratory, Department of Physics, J J Thomson Avenue, Cambridge CB3 0HE, United Kingdom}

\date{\today}

\begin{abstract}
Flexoelectricity is the polar response of an insulator to strain gradients such as bending. While the size dependence of this effect makes it weak in bulk systems in comparison to piezoelectricity, it suggests that it plays a bigger role in nanoscale systems such as thin films and nanotubes (NTs). In this paper we demonstrate using first-principles calculations that the walls of carbon nanotubes (CNTs) and transition metal dichalcogenide nanotubes (TMD NTs) are polarized in the radial direction, the strength of the polarization increasing as the size of the NT decreases. For CNTs and TMD NTs with chiral indices $(n,m)$, the radial polarization of the walls $P_R$ starts to diverge below $ C(n,m)/a = \sqrt{n^2 +nm + m^2} \sim 10$, where $C(n,m)$ is the circumference  of the NT and $a$ is the lattice constant of the 2D monolayer. For CNTs, $P_R$ drops to zero above this value but for TMD NTs there is a non-zero polarization above this value, which is an ionic rather than electronic effect. The size dependence of $P_R$ in the TMD NTs is interesting: it increases gradually and reaches a maximum of $P_R \sim 100 \ \si{C/cm^2}$ at $C(n,m)/a \sim 15$, then decreases until $C(n,m)/a \sim 10$ where it starts to diverge. Measurements of the radial strain on the chalcogen atoms with respect to the 2D monolayers shows that this polarization is the result of a significantly larger strain on the outer bonds than the inner bonds, but did not offer an explanation for the peculiar size dependence. These results suggest that while the walls of smaller CNTs and TMD NTs are polarized, the walls of larger TMD NTs are also polarized due to a difference in strain on the inner and outer atoms in the walls. This result may prove useful for the application of NTs for screening in liquid or biological systems.

\end{abstract}

\pacs{Valid PACS appear here}
\maketitle


\section{Introduction}

The polar response to an inhomogeneous strain, known as flexoelectricity, is a property of all insulators and was predicted in the late 1950s\cite{mashkevich1957electrical,tolpygo1962investigation}, with theoretical descriptions following shortly afterwards\cite{kogan1964piezoelectric,tagantsev1985theory,tagantsev1986piezoelectricity,tagantsev1991electric}. The components of the polar response $P$ to the gradient of a strain $\s$ are
\beq{polar_response}
P_i = \mu_{ijkl}\del_{l}\s_{jk}
\eec
where $\del_l \equiv \frac{\del}{\del x_l}$ and
\beq{flexo_tensor}
\mu_{ijkl} = \frac{\del P_i}{\del(\del_l\s_{jk})}
\eec
is the flexoelectric tensor. Unlike piezoelectricity, the polar response to an homogeneous strain, flexoelectricity can be observed in materials that are centrosymmetric, because a strain gradient will always locally break one or more inversion symmetries. Understanding the electromechanical properties of solids such as piezoelectric and flexoelectric responses is essential for their practical application in technology. 

Flexoelectricity is not as widely-known or applied as piezoelectricity, likely for two reasons. Firstly, it is a more complex phenomenon than piezoelectricity. Significant progress on the development of theoretical descriptions of flexoelectricity has been made in the last few decades, however\cite{PhysRevB.80.054109,PhysRevLett.105.127601,PhysRevB.84.180101,PhysRevB.88.174107,PhysRevB.88.174106,stengel2013microscopic,PhysRevB.98.075153,PhysRevB.99.085107,PhysRevX.9.021050}. Significant progress has also been made on first-principles descriptions of flexoelectricity, including density functional theory\cite{hong2010flexoelectricity} (DFT), effective modelling\cite{ponomareva2012finite}, and, perhaps most successfully, density functional perturbation theory\cite{PhysRevB.88.174106,stengel2013microscopic,PhysRevB.98.075153,PhysRevB.99.085107,PhysRevX.9.021050} (DFPT) approaches. Some aspects were not well understood, such as the distinction between surface and bulk effects\cite{PhysRevLett.105.127601,PhysRevB.90.201112}. The components of the flexoelectric tensor, i.e. the flexoelectric coefficients, are also difficult to calculate, and consistent results remained elusive\cite{zubko2007strain,PhysRevB.80.054109,hong2010flexoelectricity} until recently\cite{PhysRevB.90.201112,PhysRevB.99.085107,PhysRevX.9.021050}. The second reason is that flexoelectricity is a size dependent effect, which scales as $e/a$, where $e$ is the electron charge and $a$ is the system size\cite{tagantsev1986piezoelectricity,tagantsev1991electric,maranganti2006electromechanical,majdoub2008enhanced,size_effects}. In bulk-like systems it is negligible in comparison to piezoelectricity. 

While the second reason seems to suggest that it is not a very useful effect, it actually implies that flexoelectricity would be most prominent and have the greatest potential for applications in nanoscale systems, such as thin films and 2D materials\cite{ma2002flexoelectric,cross2006flexoelectric,ma2006flexoelectricity,
catalan2004effect,ma2008study,zubko2007strain}; it has been suggested that flexoelectricity could be utilized in electromechanical devices\cite{bhaskar2016flexoelectric}, as well as for energy and information technology\cite{catalan2011flexoelectric,lu2012mechanical,narvaez2016enhanced}. It has also been found that flexoelectricity plays a role in the bending and vibration of piezoelectric nanobeams\cite{nanobeam_1,Yan_2013}. In addition to solid devices, flexoelectricity in liquid and biological systems is an active field of research, and plays a significant role in liquid crystals and biological membranes, for example\cite{ahmadpoor2015flexoelectricity,deng2014flexoelectricity,
PETROV1996845,10.1117/12.301309,PhysRevE.72.041701,doi:10.1063/1.3358391,doi:10.1080/02678292.2011.603846}. There are already many excellent reviews in the literature, both general\cite{zubko2013flexoelectric,yudin2013fundamentals,nguyen2013nanoscale,tagantsev2016flexoelectricity,shu2019flexoelectric} and more focused ones on 2D and biological systems\cite{liu2014energy,rey2006liquid,gao2008electromechanical,mohammadi2014theory,deng2014flexoelectricity,ahmadpoor2015flexoelectricity}.

While flexoelectricity is a property of all insulators, calculations are typically restricted to cubic crystals such as strontium titanate (\chem{SrTiO_3}, STO), since the number of independent flexoelectric coefficients reduces to 3. This is unfortunate, because flexoelectricity has the most potential in low-dimensional systems, not bulk. Flexoelectric effects have been studied and observed in graphene and graphene based nanostructures\cite{PhysRevLett.102.217601,DUMITRICA2002182,PhysRevB.77.033403,zhang2010influence,kvashnin2015flexoelectricity}, as well as in transition metal dichalcogenides (TMD) monolayers\cite{shi2018flexoelectricity,han2019tunable,HAN2019471,PhysRevMaterials.3.125402,PhysRevB.99.054105}, however.

The effects of flexoelectricity in 1D structures i.e. nanotubes (NTs) is not as well known, however. In addition to carbon nanotubes (CNTs), it is also possible to fabricate transition metal dichalcogenide nanotubes (TMD NTs) from monolayers such as \chem{WS_2}\cite{tenne1992polyhedral}, \chem{MoS_2}\cite{margulis1993nested}, etc. Although not as widely studied and used as CNTs, structural and electronic properties of TMD NTs have been investigated using first-principles calculations\cite{seifert2000structure,milovsevic2007electronic,tenne2010recent,zibouche2012layers,zibouche2012layers}. However, to our knowledge, the role of flexoelectricity in TMD NTs has not been investigated. The effect of flexoelectricity on electronic and optical properties of single- and double-wall CNTs has been investigated very recently, however\cite{doi:10.1021/acs.nanolett.9b05345}.

If a NT is formed by rolling a 2D layer of finite thickness, there would be a difference in strain between the inside and outside of the wall. Hence we would expect the wall of the NT to have a finite polarization around the wall in the radial direction in response to this difference in strain. We would naturally expect this effect to occur in TMD NTs since the walls are three atoms thick. In fact, we would still expect this to occur in CNTs, but it would be a purely electronic effect in this case.

The aim of this paper is to illustrate, using first-principles DFT calculations, that the walls of both CNTs and TMD NTs in general have a radial polarization. It is difficult to measure polarization of low-dimensional structures using DFT calculations, since the Berry phase method cannot be applied. We adopt the approach of using the macroscopic potentials\cite{junquera2003first,nanosmooth,stengel2011band}, which has been successful in studying the polarization of thin films and interfaces. Normally, the electrostatic potential is obtained in the direction normal to the thin film of interface. After smoothing the rapid oscillations caused by the ions, the polarization can be estimated by measuring the electric field or voltage drop across the thin film or interface. Since the number of atoms in the walls of these NTs is small ($\leq 3$), it is not possible to smooth out the oscillations, nor is it possible to define the exact width of the wall. However, a potential drop between the inside and the outside of the wall would indicate that there is a finite electric field in the wall, and hence a polarization.

\section{Results}

\begin{figure*}[t] 
\hspace*{-0.4cm}
\centering
\includegraphics[width=\textwidth]{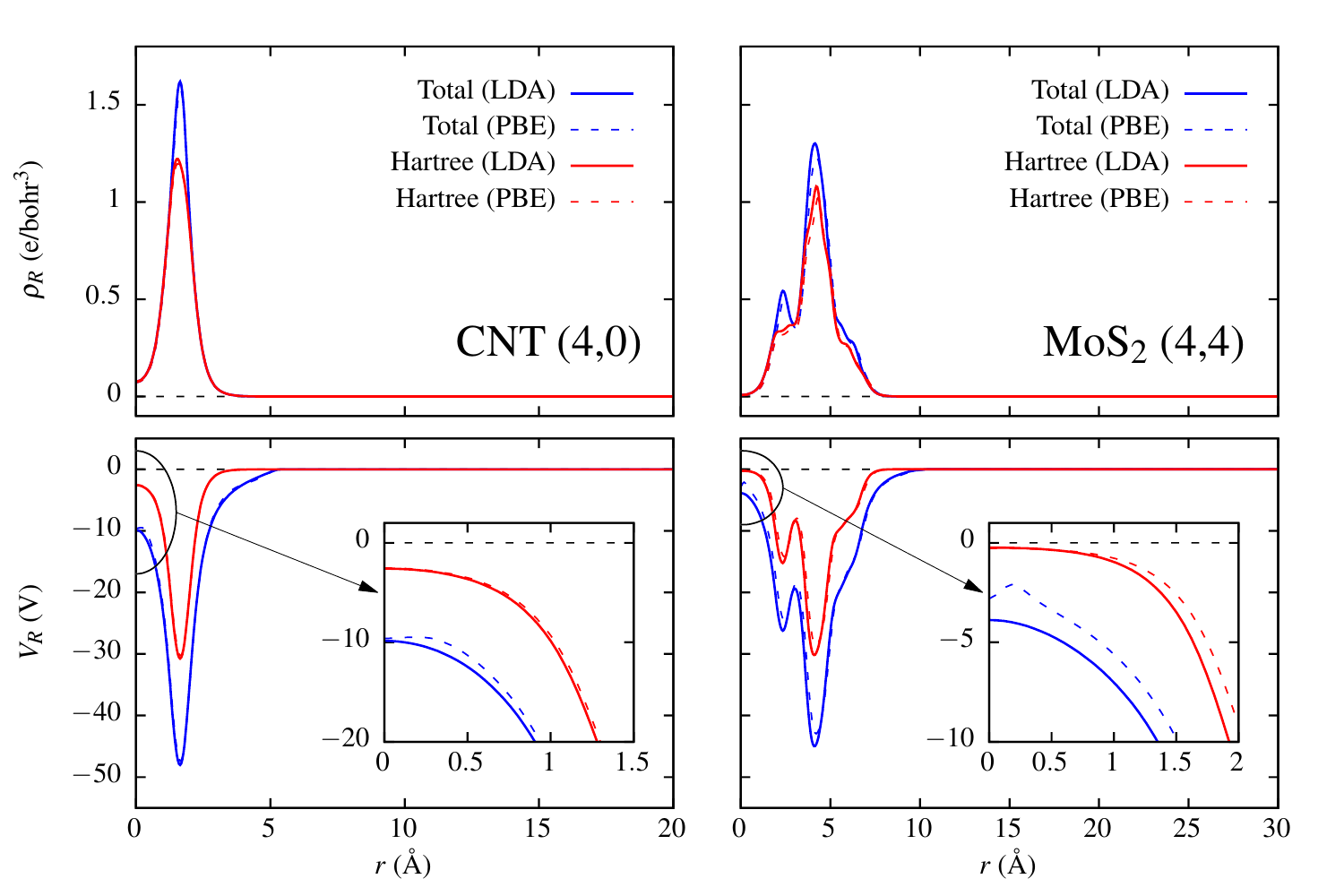}
\caption{Radial electrostatic densities (top) and potentials (bottom) for a (4,0) CNT (left) and a (4,4) \chem{MoS_2} NT (right). The red lines show the Hartree density and potential, which does not include exchange-correlation contributions and the blue lines show the total density and potential, which does. The solid lines show results obtained using the LDA functional and the dashed lines show results obtained using the PBE functional. The horizontal dashed black line indicates the value of the density and potential on the outside of the NT. The insets in the bottom plots show a close up of the potentials close to the center of the NT.}
\label{fig:potential}
\end{figure*}

First-principles DFT calculations were performed using the SIESTA code\cite{siesta} using PSML\cite{psml,javier_psml} norm-conserving\cite{norm_conserving} pseudopotentials, obtained from pseudo-dojo\cite{pseudodojo}. SIESTA employs a basis set of numerical atomic orbitals (NAOs)\cite{siesta,siesta_2}, and double-$\zeta$ polarized (DZP) orbitals were used for all calculations. The basis sets were optimised by hand, following the methodology in Ref. [\onlinecite{basis_water}]. Calculations were performed using both Perdew-Wang (PW92)\cite{pw92} and Perdew-Burke-Edwards (PBE)\cite{pbe} functionals, within the local density approximation (LDA) and generalized gradient approximation (GGA), respectively. A mesh cutoff of $800 \ \si{Ry}$ was used for the real space grid in all calculations. A Monkhorst-Pack $k$-point grid\cite{mp} of $12 \times 12 \times 1$ was used for the 2D monolayers and a grid of $1\times 1\times 12$ was used for the NTs.

Calculations were first performed for graphene, and the TMDs \chem{MoS_2}, \chem{MoSe_2}, \chem{WS_2} and \chem{WSe_2} to be used as a reference geometry for the NTs. A dipole correction\cite{dipole_correction_1,dipole_correction_2,dipole_correction_3,dipole_correction_4} was used in the vacuum region to prevent long-range interactions between periodic images\footnote{Although this is not necessary if the monolayers are correctly relaxed as they should be non-polar.}. To perform the geometry relaxations, the size of the in-plane lattice vectors $a$ was relaxed while preserving the angle between them and the in-plane atomic positions. For the TMDs, the out-of-plane atomic coordinates were also relaxed in order to obtained the height $h$ of the monolayers. The results were found to be in agreement to similar calculations in the literature\cite{computational_database_2d} and are summarised in Table \ref{table:2d}.

\begin{table}[!t]
\renewcommand*{\arraystretch}{1.25}
\setlength{\tabcolsep}{12pt}
\centering
\begin{tabular}{c | c c | c c}
\hline\hline
\multirow{2}{*}{Material} &  \multicolumn{2}{c|}{LDA} & \multicolumn{2}{c}{PBE} \\ 
&  $a$ (\AA) & $h$ (\AA) & $a$ (\AA) & $h$ (\AA) \\ \hline
Graphene & 2.468 & - & 2.475 & - \\ \hline
\chem{MoS_2} & 3.141 & 3.137 & 3.207 & 3.155 \\ \hline
\chem{MoSe_2} & 3.282 &  3.367& 3.356 & 3.393 \\ \hline
\chem{WS_2} & 3.151 & 3.140 & 3.217 & 3.163 \\ \hline
\chem{WSe_2} & 3.286 & 3.377 &  3.358 & 3.407 \\ \hline \hline
\end{tabular}
\caption{Lattice constants $a$ and heights $h$ of the 2D monolayers using both LDA and PBE functionals.}
\label{table:2d}
\end{table}

NT structures were generated from the relaxed monolayers using the c2x utility\cite{c2x}. The atomic coordinates in the plane of the circumference of the NTs, and the $c$ lattice vector were relaxed, but the atomic coordinates along the $c$ lattice vector were fixed.

A number of quantities obtained from the DFT calculations can be used to measure strain effects and estimate the polarization in the NTs. The strain energy per atom is a typical quantity used to measure such strain effects:
\beq{}
E_{\text{strain}} = \frac{E_{\text{tot}}-n_{\text{cells}}E_{\text{2D}}}{n_{\text{tot}}}
\eec
where $E_{\text{tot}}$ is the total energy of the NT, $E_{\text{2D}}$ is the energy of the monolayer, $n_{\text{tot}}$ is the number of atoms in the NT and $n_{\text{cells}} = \frac{n_{\text{tot}}}{n_{\text{2D}}}$ is the number of unit cells of the monolayer required to form the NT. It is well known that $E_{\text{strain}} \sim R^{-2}$ in general, where $R$ is the radius of the NT.

The electrostatic potentials and densities obtained from SIESTA, both with (total) and without (Hartree) exchange-correlation contributions, were converted to radial potentials using the c2x utility\cite{c2x}:
\beq{}
\begin{split}
V_R(r) &= \frac{1}{2\pi c}\int V(\vec{r})\dd z \dd \theta \\
\rho_R(r) &= \frac{1}{2\pi c}\int \rho(\vec{r})\dd z \dd \theta \\
\end{split}
\eec
where $r$ is the distance from the center of the NT. Examples of these potentials are plotted in Fig.~\ref{fig:potential} for a (4,0) CNT and a (4,4) \chem{MoS_2} NT. The internal electric field in the wall, and hence the polarization, is proportional to the potential drop $\D V_R$ across the wall:
\beq{}
\D V_R = V_R(L/2) - V_R(0)
\eeq
where $L\times L$ is the size of the unit cell in the plane of the circumference of the NT. Measuring the radius of the TMD NTs is a subtle problem. For CNTs, The standard formula 
\beq{radius}
R(n,m) = \frac{C(n,m)}{2\pi} \equiv \frac{a}{2\pi}\sqrt{n^2+nm+m^2}
\eec
is normally used, where $(n,m)$ are the chiral indices, $C(n,m)$ is the circumference and $a$ is the length of the in-plane lattice vectors of graphene (see Table \ref{table:2d}). This is exactly the same as the radius obtained from the relaxed geometry obtained from DFT calculations, except when the radius is very small. The radii are also sensitive to DFT parameters such as the exchange-correlation functional, for example. 

For TMD NTs, there are three atoms in the wall, and hence three radii: $R_{\text{S,inner}}$, $R_{\text{Mo}}$ and $R_{\text{S,outer}}$. Eq.~\eqref{radius} does not correspond to any of them. In order to estimate the polarization of the walls of the TMD NTs we also require the thickness of the wall. As with the monolayers, we will take this to be the distance between the chalcogen atoms, $R_{\text{S,outer}}-R_{\text{S,inner}}$, however as we can see from the density plots in Fig.~\ref{fig:potential} the wall extends slightly beyond the chalcogen atoms and hence this is will a lower bound on the thickness of the wall. In order to avoid confusion and for consistency across different NTs and exchange-correlation functionals we use $C(n,m)/a = \sqrt{n^2+nm+m^2}$ as a measure of the size of the NTs, as it only depends on the chiral indices. We can estimate the polarization across the walls of the TMD NTs using
\beq{pol}
P_R \sim \eo\Ep_R = \frac{\eo\D V_R}{R_{\text{S,outer}}-R_{\text{S,inner}}}
\eec
where $\Ep_R$ is the electric field across the wall. However this will be an upper bound for the reasons stated above. We can calculate the `radial' strain on the bonds with respect to the monolayers:
\beq{strain}
e_{\text{in/out}} = \frac{\Delta R_{\text{in/out}}- h/2}{h/2}
\eec
where
\beq{}
\begin{split}
\Delta R_{\text{in}} & = \left| R_{\text{S,inner}} - R_{\text{Mo}}\right| \\
\Delta R_{\text{out}} & = \left| R_{\text{S,outer}} - R_{\text{Mo}}\right|
\end{split}
\eec
and $h$ is the height of the monolayers, as in Table \ref{table:2d}.

\subsection{Carbon Nanotubes}

\begin{figure}[h] 
\hspace*{-1cm}
\centering
\includegraphics[width=\columnwidth]{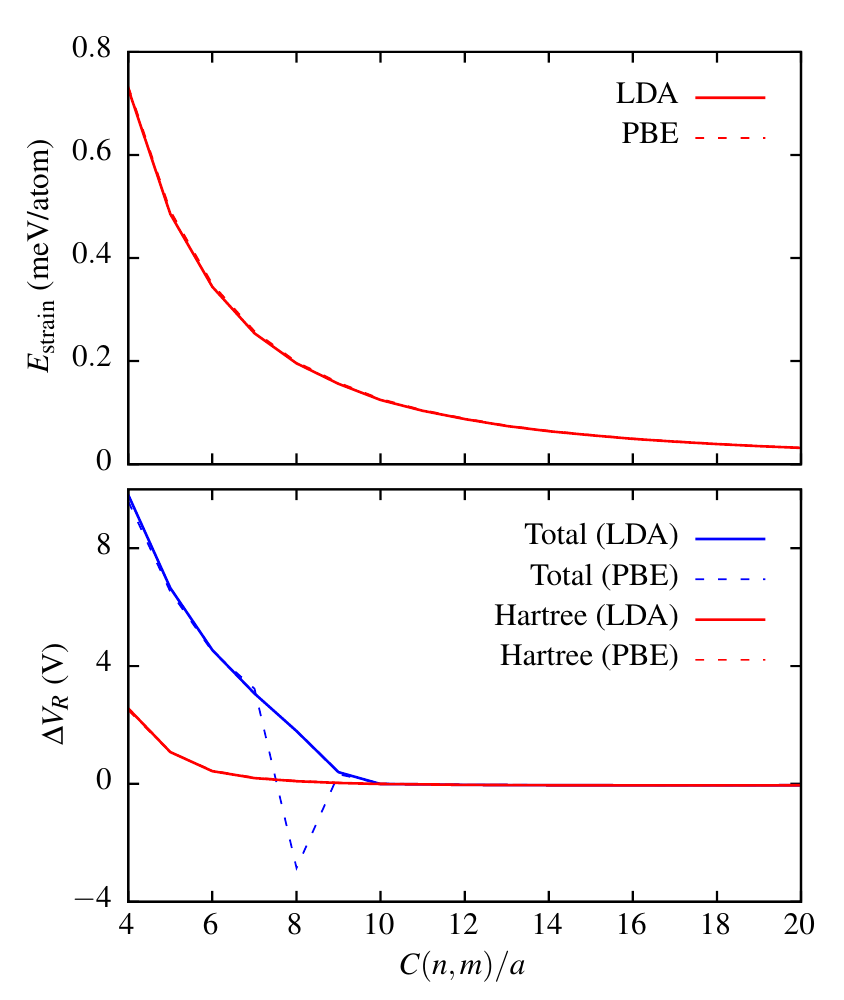}
\caption{Strain energy per atom (top) and potential drop (bottom) as a function of $C(n,m)/a$ for the zigzag CNTs. The red lines show the Hartree density and potential, which does not include exchange-correlation contributions and the blue lines show the total density and potential, which does. The solid lines show results obtained using the LDA functional and the dashed lines show results obtained using the PBE functional.}
\label{fig:cnt_energy}
\end{figure}

Geometries of zigzag $(n,0)$ CNTs were created from the graphene monolayers using c2x. The chiral index $n$ ranged from 6-20. Geometry relaxations were performed using both LDA and PBE functionals, until the forces on all atoms were less than $1 \ \si{meV/\AA}$. The electrostatic potentials and densities were then converted into radial potentials and densities using c2x. The strain energy per atom and the potential drop are plotted as a function of $C(n,m)/a$ for both functionals in Fig.~\ref{fig:cnt_energy}. We can see that the strain energy per atom is inversely proportional to size of the NTs, as expected.

The results for the potential drop are more interesting, however. The potential drop is zero until $C(n,m)/a \sim 10$, below which it begins to diverge. By plotting the drop for both Hartree and total potentials, we can see that this effect mainly comes from the exchange-correlation part of the potential. Both LDA and PBE results are in good agreement, except for the total potential in the PBE calculations, which was also found in a previous study of CNTs\cite{jwb}. For the PBE calculations, we found some unusual behaviour of the electrostatic potential at the center of the smaller NTs, which can be seen in the insets of Fig.~\ref{fig:potential}. The radial potentials exhibited strange kinks or spikes at the center of the NTs, where they should be flat. This was not observed in the LDA calculations however.

One caveat with these results is zigzag CNTs with chiral indices $(3n,0)$ should metallic (as should all armchair CNTs). This did not have an effect on the potential drop across the walls, but making any conclusions about them being polarized would not be sensible. This is not a problem for the TMD NTs however, as they are all semiconducting.

\subsection{TMD Nanotubes}

\begin{figure}[h] 
\hspace*{-1cm}
\centering
\includegraphics[width=\columnwidth]{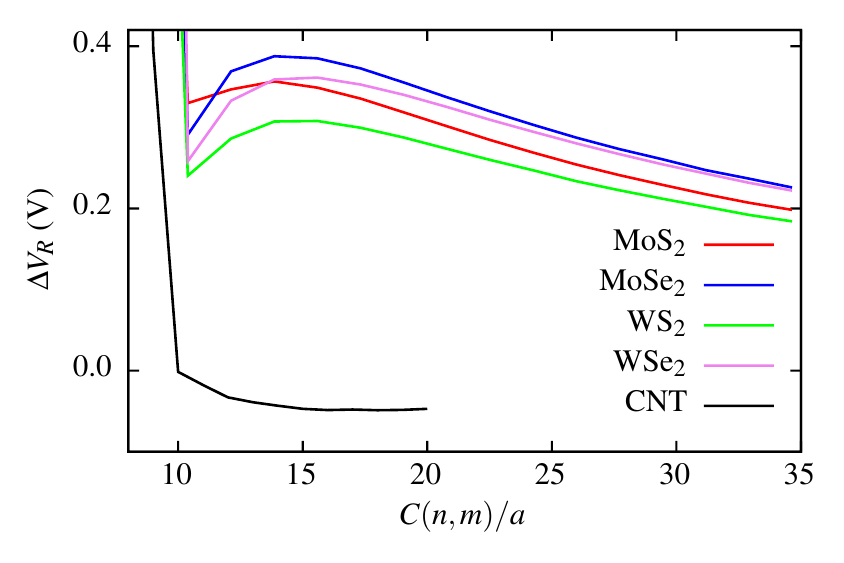}
\caption{Potential drops for the TMD NTs as a function of $C(n,m)/a$, obtained from the LDA calculations. The potential drop for the CNTs is also plotted in black for comparison.}
\label{fig:drops}
\end{figure}

Similar calculations were performed for \chem{MoS_2}, \chem{MoSe_2}, \chem{WS_2} and \chem{WSe_2} NTs, both zigzag $(n,0)$ and armchair $(n,n)$ with chiral indices $n$ ranging from 6-20 and 4-20, respectively. The strain energy per atom was found to have similar size dependence as the CNTs in all cases. The potential drops for the TMD NTs are shown in Fig.~\ref{fig:drops}

\begin{figure}[h] 
\hspace*{-1cm}
\centering
\includegraphics[width=\columnwidth]{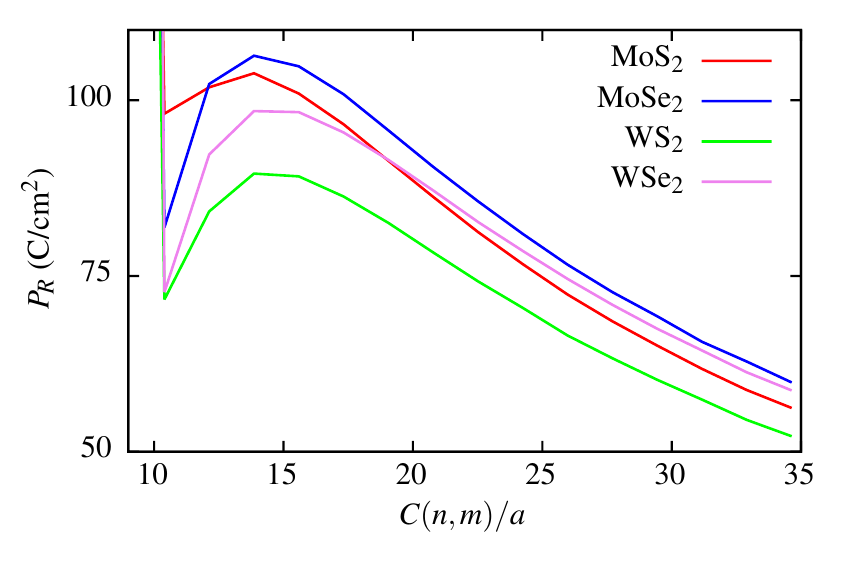}
\caption{Estimation of the radial polarization $P_R$ in the walls of the TMD NTs as a function of $C(n,m)/a$, obtained from the LDA calculations using Eq.~\eqref{pol}.}
\label{fig:pol}
\end{figure}

The potential drops of the TMD NTs also diverged below $C(n,m)/a\sim 10$, as in the case of the CNTs. With the TMD NTs however, the potential is non-zero above this value. The potential drop increases gradually as $C(n,m)/a$ decreases. There is a maximum at $C(n,m)/a \sim 15$, below which it starts to decrease, before eventually diverging at $C(n,m)/a\sim 10$. This behaviour is not observed in the CNTs, and thus we can conclude that it arises from a difference in strain on the bonds in the walls of the TMD NTs. We can use Eq.~\eqref{pol} to estimate the polarization across the wall from the potential drops in Fig.~\ref{fig:drops}. In Fig.~\ref{fig:pol} we plot the polarization in the walls, $P_R$, of the TMD NTs. First we note that, even before the potential drop diverges, the polarization across the walls is of the order of $100 \ \si{C/cm^2}$, which is not insignificant. The maximum at $C(n,m)/a\sim 15$ is interesting and unexpected behaviour. Typically we expect strain effects to decrease monotonically with NT size, such as the strain energy per atom in Fig.~\ref{fig:cnt_energy}.

\begin{figure}[h] 
\hspace*{-1cm}
\centering
\includegraphics[width=\columnwidth]{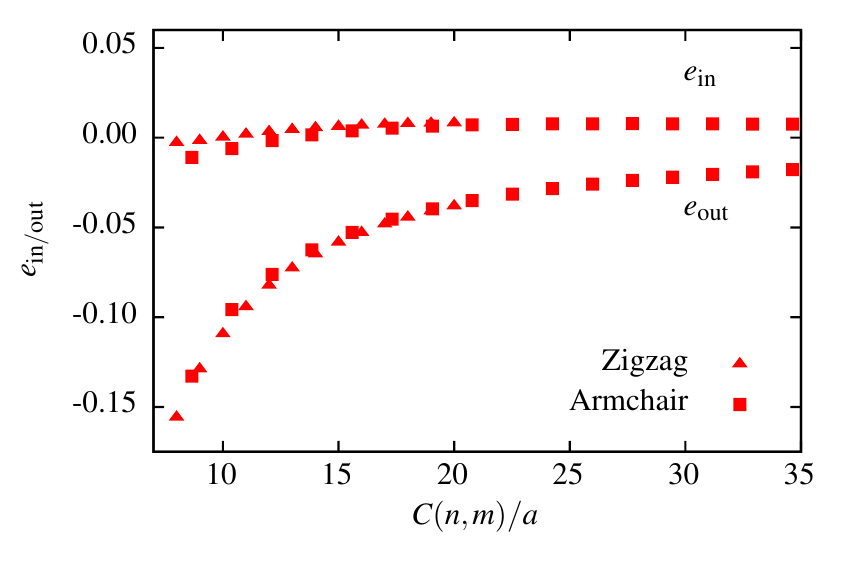}
\caption{Radial strain on the inner and outer bounds as a function of $C(n,m)/a$ for the \chem{MoS_2} NTs, obtained using the LDA calculations.}
\label{fig:mos2_strain}
\end{figure}

In Fig.~\ref{fig:mos2_strain} we plot the radial strain on the inner and outer bonds, obtained using Eq.~\eqref{strain}, for the \chem{MoS_2} NTs. From this we can see that the strain on the bonds increases monotonically as $C(n,m)/a$ decreases, which does not explain the maximum of the polarization. It does however illustrate that the polarization is a result of an inhomogeneous compression of the inner and outer bonds, the compression of the outer bond being much larger.

\section{Discussions and Conclusions}

In this paper we used first-principles DFT calculations to illustrate that radial polarization can be found in all NTs, even when the walls are a only single atom thick. We saw that the electronic polarization diverges in CNTs below $C(n,m)/a \sim 10$ by calculating the drop in the electrostatic potential across the walls of the NTs. A similar effect was observed in TMD NTs, however they displayed a finite polarization even when $C(n,m)/a > 10$. We found that polarization of the TMD NTs reaches a maximum around $C(n,m)/a \sim 15$ and decreases below before eventually diverging like the polarization of the CNTs. This behaviour was observed in all TMD NTs and is interesting and unexpected, but we are not sure what it is caused by. Plots of the radial strain on the bonds in the TMD NTs did not reveal an explanation for this maximum, but did reveal that the polarization in the walls is mainly caused by a larger compression on the outer bonds than the inner bonds.

This paper does not provide a theory of flexoelectricity in NTs, but it does show that the flexoelectric response of the walls of NTs is interesting. In fact, it is even debatable whether or not the polarization observed here should be called flexoelectric polarization. Typically one thinks of flexoelectric polarization as the response to an applied strain gradient such as bending, whereas the strain gradient here is intrinsic to the geometry of the NTs, and there is no reference state for a given NT with zero strain gradient. One could think of the parameter that controls the strain as the inverse radius, $1/R$, such that the reference state is a flat 2D monolayer with $P_R=0$, obtained when $R\to\infty$, i.e. when $1/R\to 0$. Thus, the polarization is not a parameter that is directly tunable in experiment, but rather is fixed for a given nanotube.

In any case, the behaviour of the polarization of the walls with the size of the NTs is unexpected. The total dipole moment of the NTs is still zero, but the \textit{local} polarization is non-zero. Because of this, there will be no induced dipole-dipole interactions between NTs, which is known from theory and other first-principles calculations. However, in situations where local polarization can have an effect, such as the interactions between NTs and liquids, this may be significant. It has already been seen that liquid crystals can have their flexoelectric coefficients enhanced via doping with CNTs \cite{moghadas2019flexoelectric}. Thus, this polarization could have a large influence on the behaviour water and other liquids or biomolecules inside or in the vicinity of NTs. Knowledge of this effect and its influence on liquids and biomaterials could lead to advances in technology on the nanoscale, such as nanofiltration or screening of impurities in liquids, for example\cite{lu2004finite,li2005screening}.

\section*{Acknowledgments}
The author would like to thank Dr.~Michael Rutter for useful discussions and technical support, particularly with c2x, and Kate Reidy for useful discussions. The author acknowledges support from the EPSRC Centre for Doctoral Training in Computational Methods for Materials Science under grant number EP/L015552/1. First-principles calculations were performed using resources provided by the Cambridge Service for Data Driven Discovery (CSD3) operated by the University of Cambridge Research Computing Service (http://www.csd3.cam.ac.uk/).



\begin{thebibliography}{90}%
\makeatletter
\providecommand \@ifxundefined [1]{%
 \@ifx{#1\undefined}
}%
\providecommand \@ifnum [1]{%
 \ifnum #1\expandafter \@firstoftwo
 \else \expandafter \@secondoftwo
 \fi
}%
\providecommand \@ifx [1]{%
 \ifx #1\expandafter \@firstoftwo
 \else \expandafter \@secondoftwo
 \fi
}%
\providecommand \natexlab [1]{#1}%
\providecommand \enquote  [1]{``#1''}%
\providecommand \bibnamefont  [1]{#1}%
\providecommand \bibfnamefont [1]{#1}%
\providecommand \citenamefont [1]{#1}%
\providecommand \href@noop [0]{\@secondoftwo}%
\providecommand \href [0]{\begingroup \@sanitize@url \@href}%
\providecommand \@href[1]{\@@startlink{#1}\@@href}%
\providecommand \@@href[1]{\endgroup#1\@@endlink}%
\providecommand \@sanitize@url [0]{\catcode `\\12\catcode `\$12\catcode
  `\&12\catcode `\#12\catcode `\^12\catcode `\_12\catcode `\%12\relax}%
\providecommand \@@startlink[1]{}%
\providecommand \@@endlink[0]{}%
\providecommand \url  [0]{\begingroup\@sanitize@url \@url }%
\providecommand \@url [1]{\endgroup\@href {#1}{\urlprefix }}%
\providecommand \urlprefix  [0]{URL }%
\providecommand \Eprint [0]{\href }%
\providecommand \doibase [0]{http://dx.doi.org/}%
\providecommand \selectlanguage [0]{\@gobble}%
\providecommand \bibinfo  [0]{\@secondoftwo}%
\providecommand \bibfield  [0]{\@secondoftwo}%
\providecommand \translation [1]{[#1]}%
\providecommand \BibitemOpen [0]{}%
\providecommand \bibitemStop [0]{}%
\providecommand \bibitemNoStop [0]{.\EOS\space}%
\providecommand \EOS [0]{\spacefactor3000\relax}%
\providecommand \BibitemShut  [1]{\csname bibitem#1\endcsname}%
\let\auto@bib@innerbib\@empty
\bibitem [{\citenamefont {Mashkevich}\ and\ \citenamefont
  {Tolpygo}(1957)}]{mashkevich1957electrical}%
  \BibitemOpen
  \bibfield  {author} {\bibinfo {author} {\bibfnamefont {V.}~\bibnamefont
  {Mashkevich}}\ and\ \bibinfo {author} {\bibfnamefont {K.}~\bibnamefont
  {Tolpygo}},\ }\href@noop {} {\bibfield  {journal} {\bibinfo  {journal} {Sov.
  Phys. JETP}\ }\textbf {\bibinfo {volume} {5}},\ \bibinfo {pages} {435}
  (\bibinfo {year} {1957})}\BibitemShut {NoStop}%
\bibitem [{\citenamefont {Tolpygo}(1962)}]{tolpygo1962investigation}%
  \BibitemOpen
  \bibfield  {author} {\bibinfo {author} {\bibfnamefont {K.}~\bibnamefont
  {Tolpygo}},\ }\href@noop {} {\bibfield  {journal} {\bibinfo  {journal} {Sov.
  Phys.—Solid States}\ }\textbf {\bibinfo {volume} {4}},\ \bibinfo {pages}
  {1765} (\bibinfo {year} {1962})}\BibitemShut {NoStop}%
\bibitem [{\citenamefont {Kogan}(1964)}]{kogan1964piezoelectric}%
  \BibitemOpen
  \bibfield  {author} {\bibinfo {author} {\bibfnamefont {S.~M.}\ \bibnamefont
  {Kogan}},\ }\href@noop {} {\bibfield  {journal} {\bibinfo  {journal} {Soviet
  Physics-Solid State}\ }\textbf {\bibinfo {volume} {5}},\ \bibinfo {pages}
  {2069} (\bibinfo {year} {1964})}\BibitemShut {NoStop}%
\bibitem [{\citenamefont {Tagantsev}(1985)}]{tagantsev1985theory}%
  \BibitemOpen
  \bibfield  {author} {\bibinfo {author} {\bibfnamefont {A.}~\bibnamefont
  {Tagantsev}},\ }\href@noop {} {\bibfield  {journal} {\bibinfo  {journal}
  {Zhurnal Eksperimental'noi i Teoreticheskoi Fiziki}\ }\textbf {\bibinfo
  {volume} {88}},\ \bibinfo {pages} {2108} (\bibinfo {year}
  {1985})}\BibitemShut {NoStop}%
\bibitem [{\citenamefont {Tagantsev}(1986)}]{tagantsev1986piezoelectricity}%
  \BibitemOpen
  \bibfield  {author} {\bibinfo {author} {\bibfnamefont {A.}~\bibnamefont
  {Tagantsev}},\ }\href@noop {} {\bibfield  {journal} {\bibinfo  {journal}
  {Physical Review B}\ }\textbf {\bibinfo {volume} {34}},\ \bibinfo {pages}
  {5883} (\bibinfo {year} {1986})}\BibitemShut {NoStop}%
\bibitem [{\citenamefont {Tagantsev}(1991)}]{tagantsev1991electric}%
  \BibitemOpen
  \bibfield  {author} {\bibinfo {author} {\bibfnamefont {A.~K.}\ \bibnamefont
  {Tagantsev}},\ }\href@noop {} {\bibfield  {journal} {\bibinfo  {journal}
  {Phase Transitions: A Multinational Journal}\ }\textbf {\bibinfo {volume}
  {35}},\ \bibinfo {pages} {119} (\bibinfo {year} {1991})}\BibitemShut
  {NoStop}%
\bibitem [{\citenamefont {Maranganti}\ and\ \citenamefont
  {Sharma}(2009)}]{PhysRevB.80.054109}%
  \BibitemOpen
  \bibfield  {author} {\bibinfo {author} {\bibfnamefont {R.}~\bibnamefont
  {Maranganti}}\ and\ \bibinfo {author} {\bibfnamefont {P.}~\bibnamefont
  {Sharma}},\ }\href {\doibase 10.1103/PhysRevB.80.054109} {\bibfield
  {journal} {\bibinfo  {journal} {Phys. Rev. B}\ }\textbf {\bibinfo {volume}
  {80}},\ \bibinfo {pages} {054109} (\bibinfo {year} {2009})}\BibitemShut
  {NoStop}%
\bibitem [{\citenamefont {Resta}(2010)}]{PhysRevLett.105.127601}%
  \BibitemOpen
  \bibfield  {author} {\bibinfo {author} {\bibfnamefont {R.}~\bibnamefont
  {Resta}},\ }\href {\doibase 10.1103/PhysRevLett.105.127601} {\bibfield
  {journal} {\bibinfo  {journal} {Phys. Rev. Lett.}\ }\textbf {\bibinfo
  {volume} {105}},\ \bibinfo {pages} {127601} (\bibinfo {year}
  {2010})}\BibitemShut {NoStop}%
\bibitem [{\citenamefont {Hong}\ and\ \citenamefont
  {Vanderbilt}(2011)}]{PhysRevB.84.180101}%
  \BibitemOpen
  \bibfield  {author} {\bibinfo {author} {\bibfnamefont {J.}~\bibnamefont
  {Hong}}\ and\ \bibinfo {author} {\bibfnamefont {D.}~\bibnamefont
  {Vanderbilt}},\ }\href {\doibase 10.1103/PhysRevB.84.180101} {\bibfield
  {journal} {\bibinfo  {journal} {Phys. Rev. B}\ }\textbf {\bibinfo {volume}
  {84}},\ \bibinfo {pages} {180101} (\bibinfo {year} {2011})}\BibitemShut
  {NoStop}%
\bibitem [{\citenamefont {Hong}\ and\ \citenamefont
  {Vanderbilt}(2013)}]{PhysRevB.88.174107}%
  \BibitemOpen
  \bibfield  {author} {\bibinfo {author} {\bibfnamefont {J.}~\bibnamefont
  {Hong}}\ and\ \bibinfo {author} {\bibfnamefont {D.}~\bibnamefont
  {Vanderbilt}},\ }\href {\doibase 10.1103/PhysRevB.88.174107} {\bibfield
  {journal} {\bibinfo  {journal} {Phys. Rev. B}\ }\textbf {\bibinfo {volume}
  {88}},\ \bibinfo {pages} {174107} (\bibinfo {year} {2013})}\BibitemShut
  {NoStop}%
\bibitem [{\citenamefont {Stengel}(2013{\natexlab{a}})}]{PhysRevB.88.174106}%
  \BibitemOpen
  \bibfield  {author} {\bibinfo {author} {\bibfnamefont {M.}~\bibnamefont
  {Stengel}},\ }\href {\doibase 10.1103/PhysRevB.88.174106} {\bibfield
  {journal} {\bibinfo  {journal} {Phys. Rev. B}\ }\textbf {\bibinfo {volume}
  {88}},\ \bibinfo {pages} {174106} (\bibinfo {year}
  {2013}{\natexlab{a}})}\BibitemShut {NoStop}%
\bibitem [{\citenamefont
  {Stengel}(2013{\natexlab{b}})}]{stengel2013microscopic}%
  \BibitemOpen
  \bibfield  {author} {\bibinfo {author} {\bibfnamefont {M.}~\bibnamefont
  {Stengel}},\ }\href@noop {} {\bibfield  {journal} {\bibinfo  {journal}
  {Nature communications}\ }\textbf {\bibinfo {volume} {4}},\ \bibinfo {pages}
  {1} (\bibinfo {year} {2013}{\natexlab{b}})}\BibitemShut {NoStop}%
\bibitem [{\citenamefont {Dreyer}\ \emph {et~al.}(2018)\citenamefont {Dreyer},
  \citenamefont {Stengel},\ and\ \citenamefont
  {Vanderbilt}}]{PhysRevB.98.075153}%
  \BibitemOpen
  \bibfield  {author} {\bibinfo {author} {\bibfnamefont {C.~E.}\ \bibnamefont
  {Dreyer}}, \bibinfo {author} {\bibfnamefont {M.}~\bibnamefont {Stengel}}, \
  and\ \bibinfo {author} {\bibfnamefont {D.}~\bibnamefont {Vanderbilt}},\
  }\href {\doibase 10.1103/PhysRevB.98.075153} {\bibfield  {journal} {\bibinfo
  {journal} {Phys. Rev. B}\ }\textbf {\bibinfo {volume} {98}},\ \bibinfo
  {pages} {075153} (\bibinfo {year} {2018})}\BibitemShut {NoStop}%
\bibitem [{\citenamefont {Schiaffino}\ \emph {et~al.}(2019)\citenamefont
  {Schiaffino}, \citenamefont {Dreyer}, \citenamefont {Vanderbilt},\ and\
  \citenamefont {Stengel}}]{PhysRevB.99.085107}%
  \BibitemOpen
  \bibfield  {author} {\bibinfo {author} {\bibfnamefont {A.}~\bibnamefont
  {Schiaffino}}, \bibinfo {author} {\bibfnamefont {C.~E.}\ \bibnamefont
  {Dreyer}}, \bibinfo {author} {\bibfnamefont {D.}~\bibnamefont {Vanderbilt}},
  \ and\ \bibinfo {author} {\bibfnamefont {M.}~\bibnamefont {Stengel}},\ }\href
  {\doibase 10.1103/PhysRevB.99.085107} {\bibfield  {journal} {\bibinfo
  {journal} {Phys. Rev. B}\ }\textbf {\bibinfo {volume} {99}},\ \bibinfo
  {pages} {085107} (\bibinfo {year} {2019})}\BibitemShut {NoStop}%
\bibitem [{\citenamefont {Royo}\ and\ \citenamefont
  {Stengel}(2019)}]{PhysRevX.9.021050}%
  \BibitemOpen
  \bibfield  {author} {\bibinfo {author} {\bibfnamefont {M.}~\bibnamefont
  {Royo}}\ and\ \bibinfo {author} {\bibfnamefont {M.}~\bibnamefont {Stengel}},\
  }\href {\doibase 10.1103/PhysRevX.9.021050} {\bibfield  {journal} {\bibinfo
  {journal} {Phys. Rev. X}\ }\textbf {\bibinfo {volume} {9}},\ \bibinfo {pages}
  {021050} (\bibinfo {year} {2019})}\BibitemShut {NoStop}%
\bibitem [{\citenamefont {Hong}\ \emph {et~al.}(2010)\citenamefont {Hong},
  \citenamefont {Catalan}, \citenamefont {Scott},\ and\ \citenamefont
  {Artacho}}]{hong2010flexoelectricity}%
  \BibitemOpen
  \bibfield  {author} {\bibinfo {author} {\bibfnamefont {J.}~\bibnamefont
  {Hong}}, \bibinfo {author} {\bibfnamefont {G.}~\bibnamefont {Catalan}},
  \bibinfo {author} {\bibfnamefont {J.}~\bibnamefont {Scott}}, \ and\ \bibinfo
  {author} {\bibfnamefont {E.}~\bibnamefont {Artacho}},\ }\href@noop {}
  {\bibfield  {journal} {\bibinfo  {journal} {Journal of Physics: Condensed
  Matter}\ }\textbf {\bibinfo {volume} {22}},\ \bibinfo {pages} {112201}
  (\bibinfo {year} {2010})}\BibitemShut {NoStop}%
\bibitem [{\citenamefont {Ponomareva}\ \emph {et~al.}(2012)\citenamefont
  {Ponomareva}, \citenamefont {Tagantsev},\ and\ \citenamefont
  {Bellaiche}}]{ponomareva2012finite}%
  \BibitemOpen
  \bibfield  {author} {\bibinfo {author} {\bibfnamefont {I.}~\bibnamefont
  {Ponomareva}}, \bibinfo {author} {\bibfnamefont {A.}~\bibnamefont
  {Tagantsev}}, \ and\ \bibinfo {author} {\bibfnamefont {L.}~\bibnamefont
  {Bellaiche}},\ }\href@noop {} {\bibfield  {journal} {\bibinfo  {journal}
  {Physical Review B}\ }\textbf {\bibinfo {volume} {85}},\ \bibinfo {pages}
  {104101} (\bibinfo {year} {2012})}\BibitemShut {NoStop}%
\bibitem [{\citenamefont {Stengel}(2014)}]{PhysRevB.90.201112}%
  \BibitemOpen
  \bibfield  {author} {\bibinfo {author} {\bibfnamefont {M.}~\bibnamefont
  {Stengel}},\ }\href {\doibase 10.1103/PhysRevB.90.201112} {\bibfield
  {journal} {\bibinfo  {journal} {Phys. Rev. B}\ }\textbf {\bibinfo {volume}
  {90}},\ \bibinfo {pages} {201112} (\bibinfo {year} {2014})}\BibitemShut
  {NoStop}%
\bibitem [{\citenamefont {Zubko}\ \emph {et~al.}(2007)\citenamefont {Zubko},
  \citenamefont {Catalan}, \citenamefont {Buckley}, \citenamefont {Welche},\
  and\ \citenamefont {Scott}}]{zubko2007strain}%
  \BibitemOpen
  \bibfield  {author} {\bibinfo {author} {\bibfnamefont {P.}~\bibnamefont
  {Zubko}}, \bibinfo {author} {\bibfnamefont {G.}~\bibnamefont {Catalan}},
  \bibinfo {author} {\bibfnamefont {A.}~\bibnamefont {Buckley}}, \bibinfo
  {author} {\bibfnamefont {P.}~\bibnamefont {Welche}}, \ and\ \bibinfo {author}
  {\bibfnamefont {J.}~\bibnamefont {Scott}},\ }\href@noop {} {\bibfield
  {journal} {\bibinfo  {journal} {Physical Review Letters}\ }\textbf {\bibinfo
  {volume} {99}},\ \bibinfo {pages} {167601} (\bibinfo {year}
  {2007})}\BibitemShut {NoStop}%
\bibitem [{\citenamefont {Maranganti}\ \emph {et~al.}(2006)\citenamefont
  {Maranganti}, \citenamefont {Sharma},\ and\ \citenamefont
  {Sharma}}]{maranganti2006electromechanical}%
  \BibitemOpen
  \bibfield  {author} {\bibinfo {author} {\bibfnamefont {R.}~\bibnamefont
  {Maranganti}}, \bibinfo {author} {\bibfnamefont {N.}~\bibnamefont {Sharma}},
  \ and\ \bibinfo {author} {\bibfnamefont {P.}~\bibnamefont {Sharma}},\
  }\href@noop {} {\bibfield  {journal} {\bibinfo  {journal} {Physical Review
  B}\ }\textbf {\bibinfo {volume} {74}},\ \bibinfo {pages} {014110} (\bibinfo
  {year} {2006})}\BibitemShut {NoStop}%
\bibitem [{\citenamefont {Majdoub}\ \emph {et~al.}(2008)\citenamefont
  {Majdoub}, \citenamefont {Sharma},\ and\ \citenamefont
  {Cagin}}]{majdoub2008enhanced}%
  \BibitemOpen
  \bibfield  {author} {\bibinfo {author} {\bibfnamefont {M.}~\bibnamefont
  {Majdoub}}, \bibinfo {author} {\bibfnamefont {P.}~\bibnamefont {Sharma}}, \
  and\ \bibinfo {author} {\bibfnamefont {T.}~\bibnamefont {Cagin}},\
  }\href@noop {} {\bibfield  {journal} {\bibinfo  {journal} {Physical Review
  B}\ }\textbf {\bibinfo {volume} {77}},\ \bibinfo {pages} {125424} (\bibinfo
  {year} {2008})}\BibitemShut {NoStop}%
\bibitem [{\citenamefont {Zhang}\ and\ \citenamefont
  {Jiang}(2014)}]{size_effects}%
  \BibitemOpen
  \bibfield  {author} {\bibinfo {author} {\bibfnamefont {Z.}~\bibnamefont
  {Zhang}}\ and\ \bibinfo {author} {\bibfnamefont {L.}~\bibnamefont {Jiang}},\
  }\href {\doibase 10.1063/1.4897367} {\bibfield  {journal} {\bibinfo
  {journal} {Journal of Applied Physics}\ }\textbf {\bibinfo {volume} {116}},\
  \bibinfo {pages} {134308} (\bibinfo {year} {2014})}\BibitemShut {NoStop}%
\bibitem [{\citenamefont {Ma}\ and\ \citenamefont
  {Cross}(2002)}]{ma2002flexoelectric}%
  \BibitemOpen
  \bibfield  {author} {\bibinfo {author} {\bibfnamefont {W.}~\bibnamefont
  {Ma}}\ and\ \bibinfo {author} {\bibfnamefont {L.~E.}\ \bibnamefont {Cross}},\
  }\href@noop {} {\bibfield  {journal} {\bibinfo  {journal} {Applied Physics
  Letters}\ }\textbf {\bibinfo {volume} {81}},\ \bibinfo {pages} {3440}
  (\bibinfo {year} {2002})}\BibitemShut {NoStop}%
\bibitem [{\citenamefont {Cross}(2006)}]{cross2006flexoelectric}%
  \BibitemOpen
  \bibfield  {author} {\bibinfo {author} {\bibfnamefont {L.~E.}\ \bibnamefont
  {Cross}},\ }\href@noop {} {\bibfield  {journal} {\bibinfo  {journal} {Journal
  of materials science}\ }\textbf {\bibinfo {volume} {41}},\ \bibinfo {pages}
  {53} (\bibinfo {year} {2006})}\BibitemShut {NoStop}%
\bibitem [{\citenamefont {Ma}\ and\ \citenamefont
  {Cross}(2006)}]{ma2006flexoelectricity}%
  \BibitemOpen
  \bibfield  {author} {\bibinfo {author} {\bibfnamefont {W.}~\bibnamefont
  {Ma}}\ and\ \bibinfo {author} {\bibfnamefont {L.~E.}\ \bibnamefont {Cross}},\
  }\href@noop {} {\bibfield  {journal} {\bibinfo  {journal} {Applied Physics
  Letters}\ }\textbf {\bibinfo {volume} {88}},\ \bibinfo {pages} {232902}
  (\bibinfo {year} {2006})}\BibitemShut {NoStop}%
\bibitem [{\citenamefont {Catalan}\ \emph {et~al.}(2004)\citenamefont
  {Catalan}, \citenamefont {Sinnamon},\ and\ \citenamefont
  {Gregg}}]{catalan2004effect}%
  \BibitemOpen
  \bibfield  {author} {\bibinfo {author} {\bibfnamefont {G.}~\bibnamefont
  {Catalan}}, \bibinfo {author} {\bibfnamefont {L.}~\bibnamefont {Sinnamon}}, \
  and\ \bibinfo {author} {\bibfnamefont {J.}~\bibnamefont {Gregg}},\
  }\href@noop {} {\bibfield  {journal} {\bibinfo  {journal} {Journal of
  Physics: Condensed Matter}\ }\textbf {\bibinfo {volume} {16}},\ \bibinfo
  {pages} {2253} (\bibinfo {year} {2004})}\BibitemShut {NoStop}%
\bibitem [{\citenamefont {Ma}(2008)}]{ma2008study}%
  \BibitemOpen
  \bibfield  {author} {\bibinfo {author} {\bibfnamefont {W.}~\bibnamefont
  {Ma}},\ }\href@noop {} {\bibfield  {journal} {\bibinfo  {journal} {Physica
  status Solidi (b)}\ }\textbf {\bibinfo {volume} {245}},\ \bibinfo {pages}
  {761} (\bibinfo {year} {2008})}\BibitemShut {NoStop}%
\bibitem [{\citenamefont {Bhaskar}\ \emph {et~al.}(2016)\citenamefont
  {Bhaskar}, \citenamefont {Banerjee}, \citenamefont {Abdollahi}, \citenamefont
  {Wang}, \citenamefont {Schlom}, \citenamefont {Rijnders},\ and\ \citenamefont
  {Catalan}}]{bhaskar2016flexoelectric}%
  \BibitemOpen
  \bibfield  {author} {\bibinfo {author} {\bibfnamefont {U.~K.}\ \bibnamefont
  {Bhaskar}}, \bibinfo {author} {\bibfnamefont {N.}~\bibnamefont {Banerjee}},
  \bibinfo {author} {\bibfnamefont {A.}~\bibnamefont {Abdollahi}}, \bibinfo
  {author} {\bibfnamefont {Z.}~\bibnamefont {Wang}}, \bibinfo {author}
  {\bibfnamefont {D.~G.}\ \bibnamefont {Schlom}}, \bibinfo {author}
  {\bibfnamefont {G.}~\bibnamefont {Rijnders}}, \ and\ \bibinfo {author}
  {\bibfnamefont {G.}~\bibnamefont {Catalan}},\ }\href@noop {} {\bibfield
  {journal} {\bibinfo  {journal} {Nature nanotechnology}\ }\textbf {\bibinfo
  {volume} {11}},\ \bibinfo {pages} {263} (\bibinfo {year} {2016})}\BibitemShut
  {NoStop}%
\bibitem [{\citenamefont {Catalan}\ \emph {et~al.}(2011)\citenamefont
  {Catalan}, \citenamefont {Lubk}, \citenamefont {Vlooswijk}, \citenamefont
  {Snoeck}, \citenamefont {Magen}, \citenamefont {Janssens}, \citenamefont
  {Rispens}, \citenamefont {Rijnders}, \citenamefont {Blank},\ and\
  \citenamefont {Noheda}}]{catalan2011flexoelectric}%
  \BibitemOpen
  \bibfield  {author} {\bibinfo {author} {\bibfnamefont {G.}~\bibnamefont
  {Catalan}}, \bibinfo {author} {\bibfnamefont {A.}~\bibnamefont {Lubk}},
  \bibinfo {author} {\bibfnamefont {A.}~\bibnamefont {Vlooswijk}}, \bibinfo
  {author} {\bibfnamefont {E.}~\bibnamefont {Snoeck}}, \bibinfo {author}
  {\bibfnamefont {C.}~\bibnamefont {Magen}}, \bibinfo {author} {\bibfnamefont
  {A.}~\bibnamefont {Janssens}}, \bibinfo {author} {\bibfnamefont
  {G.}~\bibnamefont {Rispens}}, \bibinfo {author} {\bibfnamefont
  {G.}~\bibnamefont {Rijnders}}, \bibinfo {author} {\bibfnamefont {D.~H.}\
  \bibnamefont {Blank}}, \ and\ \bibinfo {author} {\bibfnamefont
  {B.}~\bibnamefont {Noheda}},\ }\href@noop {} {\bibfield  {journal} {\bibinfo
  {journal} {Nature materials}\ }\textbf {\bibinfo {volume} {10}},\ \bibinfo
  {pages} {963} (\bibinfo {year} {2011})}\BibitemShut {NoStop}%
\bibitem [{\citenamefont {Lu}\ \emph {et~al.}(2012)\citenamefont {Lu},
  \citenamefont {Bark}, \citenamefont {De~Los~Ojos}, \citenamefont {Alcala},
  \citenamefont {Eom}, \citenamefont {Catalan},\ and\ \citenamefont
  {Gruverman}}]{lu2012mechanical}%
  \BibitemOpen
  \bibfield  {author} {\bibinfo {author} {\bibfnamefont {H.}~\bibnamefont
  {Lu}}, \bibinfo {author} {\bibfnamefont {C.-W.}\ \bibnamefont {Bark}},
  \bibinfo {author} {\bibfnamefont {D.~E.}\ \bibnamefont {De~Los~Ojos}},
  \bibinfo {author} {\bibfnamefont {J.}~\bibnamefont {Alcala}}, \bibinfo
  {author} {\bibfnamefont {C.-B.}\ \bibnamefont {Eom}}, \bibinfo {author}
  {\bibfnamefont {G.}~\bibnamefont {Catalan}}, \ and\ \bibinfo {author}
  {\bibfnamefont {A.}~\bibnamefont {Gruverman}},\ }\href@noop {} {\bibfield
  {journal} {\bibinfo  {journal} {Science}\ }\textbf {\bibinfo {volume}
  {336}},\ \bibinfo {pages} {59} (\bibinfo {year} {2012})}\BibitemShut
  {NoStop}%
\bibitem [{\citenamefont {Narvaez}\ \emph {et~al.}(2016)\citenamefont
  {Narvaez}, \citenamefont {Vasquez-Sancho},\ and\ \citenamefont
  {Catalan}}]{narvaez2016enhanced}%
  \BibitemOpen
  \bibfield  {author} {\bibinfo {author} {\bibfnamefont {J.}~\bibnamefont
  {Narvaez}}, \bibinfo {author} {\bibfnamefont {F.}~\bibnamefont
  {Vasquez-Sancho}}, \ and\ \bibinfo {author} {\bibfnamefont {G.}~\bibnamefont
  {Catalan}},\ }\href@noop {} {\bibfield  {journal} {\bibinfo  {journal}
  {Nature}\ }\textbf {\bibinfo {volume} {538}},\ \bibinfo {pages} {219}
  (\bibinfo {year} {2016})}\BibitemShut {NoStop}%
\bibitem [{\citenamefont {Yan}\ and\ \citenamefont
  {Jiang}(2013{\natexlab{a}})}]{nanobeam_1}%
  \BibitemOpen
  \bibfield  {author} {\bibinfo {author} {\bibfnamefont {Z.}~\bibnamefont
  {Yan}}\ and\ \bibinfo {author} {\bibfnamefont {L.~Y.}\ \bibnamefont
  {Jiang}},\ }\href {\doibase 10.1063/1.4804949} {\bibfield  {journal}
  {\bibinfo  {journal} {Journal of Applied Physics}\ }\textbf {\bibinfo
  {volume} {113}},\ \bibinfo {pages} {194102} (\bibinfo {year}
  {2013}{\natexlab{a}})}\BibitemShut {NoStop}%
\bibitem [{\citenamefont {Yan}\ and\ \citenamefont
  {Jiang}(2013{\natexlab{b}})}]{Yan_2013}%
  \BibitemOpen
  \bibfield  {author} {\bibinfo {author} {\bibfnamefont {Z.}~\bibnamefont
  {Yan}}\ and\ \bibinfo {author} {\bibfnamefont {L.}~\bibnamefont {Jiang}},\
  }\href {\doibase 10.1088/0022-3727/46/35/355502} {\bibfield  {journal}
  {\bibinfo  {journal} {Journal of Physics D: Applied Physics}\ }\textbf
  {\bibinfo {volume} {46}},\ \bibinfo {pages} {355502} (\bibinfo {year}
  {2013}{\natexlab{b}})}\BibitemShut {NoStop}%
\bibitem [{\citenamefont {Ahmadpoor}\ and\ \citenamefont
  {Sharma}(2015)}]{ahmadpoor2015flexoelectricity}%
  \BibitemOpen
  \bibfield  {author} {\bibinfo {author} {\bibfnamefont {F.}~\bibnamefont
  {Ahmadpoor}}\ and\ \bibinfo {author} {\bibfnamefont {P.}~\bibnamefont
  {Sharma}},\ }\href@noop {} {\bibfield  {journal} {\bibinfo  {journal}
  {Nanoscale}\ }\textbf {\bibinfo {volume} {7}},\ \bibinfo {pages} {16555}
  (\bibinfo {year} {2015})}\BibitemShut {NoStop}%
\bibitem [{\citenamefont {Deng}\ \emph {et~al.}(2014)\citenamefont {Deng},
  \citenamefont {Liu},\ and\ \citenamefont
  {Sharma}}]{deng2014flexoelectricity}%
  \BibitemOpen
  \bibfield  {author} {\bibinfo {author} {\bibfnamefont {Q.}~\bibnamefont
  {Deng}}, \bibinfo {author} {\bibfnamefont {L.}~\bibnamefont {Liu}}, \ and\
  \bibinfo {author} {\bibfnamefont {P.}~\bibnamefont {Sharma}},\ }\href@noop {}
  {\bibfield  {journal} {\bibinfo  {journal} {Journal of the Mechanics and
  Physics of Solids}\ }\textbf {\bibinfo {volume} {62}},\ \bibinfo {pages}
  {209} (\bibinfo {year} {2014})}\BibitemShut {NoStop}%
\bibitem [{\citenamefont {Petrov}\ \emph {et~al.}(1996)\citenamefont {Petrov},
  \citenamefont {Spassova},\ and\ \citenamefont {Fendler}}]{PETROV1996845}%
  \BibitemOpen
  \bibfield  {author} {\bibinfo {author} {\bibfnamefont {A.}~\bibnamefont
  {Petrov}}, \bibinfo {author} {\bibfnamefont {M.}~\bibnamefont {Spassova}}, \
  and\ \bibinfo {author} {\bibfnamefont {J.}~\bibnamefont {Fendler}},\ }\href
  {\doibase https://doi.org/10.1016/S0040-6090(95)08461-4} {\bibfield
  {journal} {\bibinfo  {journal} {Thin Solid Films}\ }\textbf {\bibinfo
  {volume} {284-285}},\ \bibinfo {pages} {845 } (\bibinfo {year} {1996})},\
  \bibinfo {note} {seventh International Conference on Organized Molecular
  Films}\BibitemShut {NoStop}%
\bibitem [{\citenamefont {Petrov}(1998)}]{10.1117/12.301309}%
  \BibitemOpen
  \bibfield  {author} {\bibinfo {author} {\bibfnamefont {A.~G.}\ \bibnamefont
  {Petrov}},\ }in\ \href {\doibase 10.1117/12.301309} {\emph {\bibinfo
  {booktitle} {Liquid Crystals: Chemistry and Structure}}},\ Vol.\ \bibinfo
  {volume} {3319},\ \bibinfo {editor} {edited by\ \bibinfo {editor}
  {\bibfnamefont {M.}~\bibnamefont {Tykarska}}, \bibinfo {editor}
  {\bibfnamefont {R.~S.}\ \bibnamefont {Dabrowski}}, \ and\ \bibinfo {editor}
  {\bibfnamefont {J.}~\bibnamefont {Zielinski}}},\ \bibinfo {organization}
  {International Society for Optics and Photonics}\ (\bibinfo  {publisher}
  {SPIE},\ \bibinfo {year} {1998})\ pp.\ \bibinfo {pages} {306 --
  318}\BibitemShut {NoStop}%
\bibitem [{\citenamefont {Kuczy\ifmmode~\acute{n}\else \'{n}\fi{}ski}\ and\
  \citenamefont {Hoffmann}(2005)}]{PhysRevE.72.041701}%
  \BibitemOpen
  \bibfield  {author} {\bibinfo {author} {\bibfnamefont {W.}~\bibnamefont
  {Kuczy\ifmmode~\acute{n}\else \'{n}\fi{}ski}}\ and\ \bibinfo {author}
  {\bibfnamefont {J.}~\bibnamefont {Hoffmann}},\ }\href {\doibase
  10.1103/PhysRevE.72.041701} {\bibfield  {journal} {\bibinfo  {journal} {Phys.
  Rev. E}\ }\textbf {\bibinfo {volume} {72}},\ \bibinfo {pages} {041701}
  (\bibinfo {year} {2005})}\BibitemShut {NoStop}%
\bibitem [{\citenamefont {Harden}\ \emph {et~al.}(2010)\citenamefont {Harden},
  \citenamefont {Chambers}, \citenamefont {Verduzco}, \citenamefont {Luchette},
  \citenamefont {Gleeson}, \citenamefont {Sprunt},\ and\ \citenamefont
  {Jákli}}]{doi:10.1063/1.3358391}%
  \BibitemOpen
  \bibfield  {author} {\bibinfo {author} {\bibfnamefont {J.}~\bibnamefont
  {Harden}}, \bibinfo {author} {\bibfnamefont {M.}~\bibnamefont {Chambers}},
  \bibinfo {author} {\bibfnamefont {R.}~\bibnamefont {Verduzco}}, \bibinfo
  {author} {\bibfnamefont {P.}~\bibnamefont {Luchette}}, \bibinfo {author}
  {\bibfnamefont {J.~T.}\ \bibnamefont {Gleeson}}, \bibinfo {author}
  {\bibfnamefont {S.}~\bibnamefont {Sprunt}}, \ and\ \bibinfo {author}
  {\bibfnamefont {A.}~\bibnamefont {Jákli}},\ }\href {\doibase
  10.1063/1.3358391} {\bibfield  {journal} {\bibinfo  {journal} {Applied
  Physics Letters}\ }\textbf {\bibinfo {volume} {96}},\ \bibinfo {pages}
  {102907} (\bibinfo {year} {2010})}\BibitemShut {NoStop}%
\bibitem [{\citenamefont {Jewell}(2011)}]{doi:10.1080/02678292.2011.603846}%
  \BibitemOpen
  \bibfield  {author} {\bibinfo {author} {\bibfnamefont {S.}~\bibnamefont
  {Jewell}},\ }\href {\doibase 10.1080/02678292.2011.603846} {\bibfield
  {journal} {\bibinfo  {journal} {Liquid Crystals}\ }\textbf {\bibinfo {volume}
  {38}},\ \bibinfo {pages} {1699} (\bibinfo {year} {2011})}\BibitemShut
  {NoStop}%
\bibitem [{\citenamefont {Zubko}\ \emph {et~al.}(2013)\citenamefont {Zubko},
  \citenamefont {Catalan},\ and\ \citenamefont
  {Tagantsev}}]{zubko2013flexoelectric}%
  \BibitemOpen
  \bibfield  {author} {\bibinfo {author} {\bibfnamefont {P.}~\bibnamefont
  {Zubko}}, \bibinfo {author} {\bibfnamefont {G.}~\bibnamefont {Catalan}}, \
  and\ \bibinfo {author} {\bibfnamefont {A.~K.}\ \bibnamefont {Tagantsev}},\
  }\href@noop {} {\bibfield  {journal} {\bibinfo  {journal} {Annual Review of
  Materials Research}\ }\textbf {\bibinfo {volume} {43}} (\bibinfo {year}
  {2013})}\BibitemShut {NoStop}%
\bibitem [{\citenamefont {Yudin}\ and\ \citenamefont
  {Tagantsev}(2013)}]{yudin2013fundamentals}%
  \BibitemOpen
  \bibfield  {author} {\bibinfo {author} {\bibfnamefont {P.}~\bibnamefont
  {Yudin}}\ and\ \bibinfo {author} {\bibfnamefont {A.}~\bibnamefont
  {Tagantsev}},\ }\href@noop {} {\bibfield  {journal} {\bibinfo  {journal}
  {Nanotechnology}\ }\textbf {\bibinfo {volume} {24}},\ \bibinfo {pages}
  {432001} (\bibinfo {year} {2013})}\BibitemShut {NoStop}%
\bibitem [{\citenamefont {Nguyen}\ \emph {et~al.}(2013)\citenamefont {Nguyen},
  \citenamefont {Mao}, \citenamefont {Yeh}, \citenamefont {Purohit},\ and\
  \citenamefont {McAlpine}}]{nguyen2013nanoscale}%
  \BibitemOpen
  \bibfield  {author} {\bibinfo {author} {\bibfnamefont {T.~D.}\ \bibnamefont
  {Nguyen}}, \bibinfo {author} {\bibfnamefont {S.}~\bibnamefont {Mao}},
  \bibinfo {author} {\bibfnamefont {Y.-W.}\ \bibnamefont {Yeh}}, \bibinfo
  {author} {\bibfnamefont {P.~K.}\ \bibnamefont {Purohit}}, \ and\ \bibinfo
  {author} {\bibfnamefont {M.~C.}\ \bibnamefont {McAlpine}},\ }\href@noop {}
  {\bibfield  {journal} {\bibinfo  {journal} {Advanced Materials}\ }\textbf
  {\bibinfo {volume} {25}},\ \bibinfo {pages} {946} (\bibinfo {year}
  {2013})}\BibitemShut {NoStop}%
\bibitem [{\citenamefont {Tagantsev}\ \emph {et~al.}(2016)\citenamefont
  {Tagantsev}, \citenamefont {Yudin},\ and\ \citenamefont
  {Tagantsev}}]{tagantsev2016flexoelectricity}%
  \BibitemOpen
  \bibfield  {author} {\bibinfo {author} {\bibfnamefont {A.~K.}\ \bibnamefont
  {Tagantsev}}, \bibinfo {author} {\bibfnamefont {P.~V.}\ \bibnamefont
  {Yudin}}, \ and\ \bibinfo {author} {\bibfnamefont {A.~K.}\ \bibnamefont
  {Tagantsev}},\ }\href@noop {} {\emph {\bibinfo {title} {Flexoelectricity in
  solids: from theory to applications}}}\ (\bibinfo  {publisher} {World
  Scientific Publishing Company},\ \bibinfo {year} {2016})\BibitemShut
  {NoStop}%
\bibitem [{\citenamefont {Shu}\ \emph {et~al.}(2019)\citenamefont {Shu},
  \citenamefont {Liang}, \citenamefont {Rao}, \citenamefont {Fei},
  \citenamefont {Ke},\ and\ \citenamefont {Wang}}]{shu2019flexoelectric}%
  \BibitemOpen
  \bibfield  {author} {\bibinfo {author} {\bibfnamefont {L.}~\bibnamefont
  {Shu}}, \bibinfo {author} {\bibfnamefont {R.}~\bibnamefont {Liang}}, \bibinfo
  {author} {\bibfnamefont {Z.}~\bibnamefont {Rao}}, \bibinfo {author}
  {\bibfnamefont {L.}~\bibnamefont {Fei}}, \bibinfo {author} {\bibfnamefont
  {S.}~\bibnamefont {Ke}}, \ and\ \bibinfo {author} {\bibfnamefont
  {Y.}~\bibnamefont {Wang}},\ }\href@noop {} {\bibfield  {journal} {\bibinfo
  {journal} {Journal of Advanced Ceramics}\ ,\ \bibinfo {pages} {1}} (\bibinfo
  {year} {2019})}\BibitemShut {NoStop}%
\bibitem [{\citenamefont {Liu}(2014)}]{liu2014energy}%
  \BibitemOpen
  \bibfield  {author} {\bibinfo {author} {\bibfnamefont {L.}~\bibnamefont
  {Liu}},\ }\href@noop {} {\bibfield  {journal} {\bibinfo  {journal} {Journal
  of the Mechanics and Physics of Solids}\ }\textbf {\bibinfo {volume} {63}},\
  \bibinfo {pages} {451} (\bibinfo {year} {2014})}\BibitemShut {NoStop}%
\bibitem [{\citenamefont {Rey}(2006)}]{rey2006liquid}%
  \BibitemOpen
  \bibfield  {author} {\bibinfo {author} {\bibfnamefont {A.~D.}\ \bibnamefont
  {Rey}},\ }\href@noop {} {\bibfield  {journal} {\bibinfo  {journal} {Physical
  Review E}\ }\textbf {\bibinfo {volume} {74}},\ \bibinfo {pages} {011710}
  (\bibinfo {year} {2006})}\BibitemShut {NoStop}%
\bibitem [{\citenamefont {Gao}\ \emph {et~al.}(2008)\citenamefont {Gao},
  \citenamefont {Feng}, \citenamefont {Yin},\ and\ \citenamefont
  {Gao}}]{gao2008electromechanical}%
  \BibitemOpen
  \bibfield  {author} {\bibinfo {author} {\bibfnamefont {L.-T.}\ \bibnamefont
  {Gao}}, \bibinfo {author} {\bibfnamefont {X.-Q.}\ \bibnamefont {Feng}},
  \bibinfo {author} {\bibfnamefont {Y.-J.}\ \bibnamefont {Yin}}, \ and\
  \bibinfo {author} {\bibfnamefont {H.}~\bibnamefont {Gao}},\ }\href@noop {}
  {\bibfield  {journal} {\bibinfo  {journal} {Journal of the Mechanics and
  Physics of Solids}\ }\textbf {\bibinfo {volume} {56}},\ \bibinfo {pages}
  {2844} (\bibinfo {year} {2008})}\BibitemShut {NoStop}%
\bibitem [{\citenamefont {Mohammadi}\ \emph {et~al.}(2014)\citenamefont
  {Mohammadi}, \citenamefont {Liu},\ and\ \citenamefont
  {Sharma}}]{mohammadi2014theory}%
  \BibitemOpen
  \bibfield  {author} {\bibinfo {author} {\bibfnamefont {P.}~\bibnamefont
  {Mohammadi}}, \bibinfo {author} {\bibfnamefont {L.}~\bibnamefont {Liu}}, \
  and\ \bibinfo {author} {\bibfnamefont {P.}~\bibnamefont {Sharma}},\
  }\href@noop {} {\bibfield  {journal} {\bibinfo  {journal} {Journal of Applied
  Mechanics}\ }\textbf {\bibinfo {volume} {81}} (\bibinfo {year}
  {2014})}\BibitemShut {NoStop}%
\bibitem [{\citenamefont {Naumov}\ \emph {et~al.}(2009)\citenamefont {Naumov},
  \citenamefont {Bratkovsky},\ and\ \citenamefont
  {Ranjan}}]{PhysRevLett.102.217601}%
  \BibitemOpen
  \bibfield  {author} {\bibinfo {author} {\bibfnamefont {I.}~\bibnamefont
  {Naumov}}, \bibinfo {author} {\bibfnamefont {A.~M.}\ \bibnamefont
  {Bratkovsky}}, \ and\ \bibinfo {author} {\bibfnamefont {V.}~\bibnamefont
  {Ranjan}},\ }\href {\doibase 10.1103/PhysRevLett.102.217601} {\bibfield
  {journal} {\bibinfo  {journal} {Phys. Rev. Lett.}\ }\textbf {\bibinfo
  {volume} {102}},\ \bibinfo {pages} {217601} (\bibinfo {year}
  {2009})}\BibitemShut {NoStop}%
\bibitem [{\citenamefont {Dumitrică}\ \emph {et~al.}(2002)\citenamefont
  {Dumitrică}, \citenamefont {Landis},\ and\ \citenamefont
  {Yakobson}}]{DUMITRICA2002182}%
  \BibitemOpen
  \bibfield  {author} {\bibinfo {author} {\bibfnamefont {T.}~\bibnamefont
  {Dumitrică}}, \bibinfo {author} {\bibfnamefont {C.~M.}\ \bibnamefont
  {Landis}}, \ and\ \bibinfo {author} {\bibfnamefont {B.~I.}\ \bibnamefont
  {Yakobson}},\ }\href {\doibase https://doi.org/10.1016/S0009-2614(02)00820-5}
  {\bibfield  {journal} {\bibinfo  {journal} {Chemical Physics Letters}\
  }\textbf {\bibinfo {volume} {360}},\ \bibinfo {pages} {182 } (\bibinfo {year}
  {2002})}\BibitemShut {NoStop}%
\bibitem [{\citenamefont {Kalinin}\ and\ \citenamefont
  {Meunier}(2008)}]{PhysRevB.77.033403}%
  \BibitemOpen
  \bibfield  {author} {\bibinfo {author} {\bibfnamefont {S.~V.}\ \bibnamefont
  {Kalinin}}\ and\ \bibinfo {author} {\bibfnamefont {V.}~\bibnamefont
  {Meunier}},\ }\href {\doibase 10.1103/PhysRevB.77.033403} {\bibfield
  {journal} {\bibinfo  {journal} {Phys. Rev. B}\ }\textbf {\bibinfo {volume}
  {77}},\ \bibinfo {pages} {033403} (\bibinfo {year} {2008})}\BibitemShut
  {NoStop}%
\bibitem [{\citenamefont {Zhang}\ \emph {et~al.}(2010)\citenamefont {Zhang},
  \citenamefont {Ong},\ and\ \citenamefont {Wu}}]{zhang2010influence}%
  \BibitemOpen
  \bibfield  {author} {\bibinfo {author} {\bibfnamefont {J.}~\bibnamefont
  {Zhang}}, \bibinfo {author} {\bibfnamefont {K.~P.}\ \bibnamefont {Ong}}, \
  and\ \bibinfo {author} {\bibfnamefont {P.}~\bibnamefont {Wu}},\ }\href@noop
  {} {\bibfield  {journal} {\bibinfo  {journal} {The Journal of Physical
  Chemistry C}\ }\textbf {\bibinfo {volume} {114}},\ \bibinfo {pages} {12749}
  (\bibinfo {year} {2010})}\BibitemShut {NoStop}%
\bibitem [{\citenamefont {Kvashnin}\ \emph {et~al.}(2015)\citenamefont
  {Kvashnin}, \citenamefont {Sorokin},\ and\ \citenamefont
  {Yakobson}}]{kvashnin2015flexoelectricity}%
  \BibitemOpen
  \bibfield  {author} {\bibinfo {author} {\bibfnamefont {A.~G.}\ \bibnamefont
  {Kvashnin}}, \bibinfo {author} {\bibfnamefont {P.~B.}\ \bibnamefont
  {Sorokin}}, \ and\ \bibinfo {author} {\bibfnamefont {B.~I.}\ \bibnamefont
  {Yakobson}},\ }\href@noop {} {\bibfield  {journal} {\bibinfo  {journal} {The
  journal of physical chemistry letters}\ }\textbf {\bibinfo {volume} {6}},\
  \bibinfo {pages} {2740} (\bibinfo {year} {2015})}\BibitemShut {NoStop}%
\bibitem [{\citenamefont {Shi}\ \emph {et~al.}(2018)\citenamefont {Shi},
  \citenamefont {Guo}, \citenamefont {Zhang},\ and\ \citenamefont
  {Guo}}]{shi2018flexoelectricity}%
  \BibitemOpen
  \bibfield  {author} {\bibinfo {author} {\bibfnamefont {W.}~\bibnamefont
  {Shi}}, \bibinfo {author} {\bibfnamefont {Y.}~\bibnamefont {Guo}}, \bibinfo
  {author} {\bibfnamefont {Z.}~\bibnamefont {Zhang}}, \ and\ \bibinfo {author}
  {\bibfnamefont {W.}~\bibnamefont {Guo}},\ }\href@noop {} {\bibfield
  {journal} {\bibinfo  {journal} {The journal of physical chemistry letters}\
  }\textbf {\bibinfo {volume} {9}},\ \bibinfo {pages} {6841} (\bibinfo {year}
  {2018})}\BibitemShut {NoStop}%
\bibitem [{\citenamefont {Han}\ \emph {et~al.}(2019{\natexlab{a}})\citenamefont
  {Han}, \citenamefont {Kim}, \citenamefont {Jang}, \citenamefont {Lim},
  \citenamefont {Kim}, \citenamefont {Chang}, \citenamefont {Song},
  \citenamefont {Lee}, \citenamefont {Lim}, \citenamefont {An} \emph
  {et~al.}}]{han2019tunable}%
  \BibitemOpen
  \bibfield  {author} {\bibinfo {author} {\bibfnamefont {J.~K.}\ \bibnamefont
  {Han}}, \bibinfo {author} {\bibfnamefont {S.}~\bibnamefont {Kim}}, \bibinfo
  {author} {\bibfnamefont {S.}~\bibnamefont {Jang}}, \bibinfo {author}
  {\bibfnamefont {Y.~R.}\ \bibnamefont {Lim}}, \bibinfo {author} {\bibfnamefont
  {S.-W.}\ \bibnamefont {Kim}}, \bibinfo {author} {\bibfnamefont
  {H.}~\bibnamefont {Chang}}, \bibinfo {author} {\bibfnamefont
  {W.}~\bibnamefont {Song}}, \bibinfo {author} {\bibfnamefont {S.~S.}\
  \bibnamefont {Lee}}, \bibinfo {author} {\bibfnamefont {J.}~\bibnamefont
  {Lim}}, \bibinfo {author} {\bibfnamefont {K.-S.}\ \bibnamefont {An}},  \emph
  {et~al.},\ }\href@noop {} {\bibfield  {journal} {\bibinfo  {journal} {Nano
  Energy}\ }\textbf {\bibinfo {volume} {61}},\ \bibinfo {pages} {471} (\bibinfo
  {year} {2019}{\natexlab{a}})}\BibitemShut {NoStop}%
\bibitem [{\citenamefont {Han}\ \emph {et~al.}(2019{\natexlab{b}})\citenamefont
  {Han}, \citenamefont {Kim}, \citenamefont {Jang}, \citenamefont {Lim},
  \citenamefont {Kim}, \citenamefont {Chang}, \citenamefont {Song},
  \citenamefont {Lee}, \citenamefont {Lim}, \citenamefont {An},\ and\
  \citenamefont {Myung}}]{HAN2019471}%
  \BibitemOpen
  \bibfield  {author} {\bibinfo {author} {\bibfnamefont {J.~K.}\ \bibnamefont
  {Han}}, \bibinfo {author} {\bibfnamefont {S.}~\bibnamefont {Kim}}, \bibinfo
  {author} {\bibfnamefont {S.}~\bibnamefont {Jang}}, \bibinfo {author}
  {\bibfnamefont {Y.~R.}\ \bibnamefont {Lim}}, \bibinfo {author} {\bibfnamefont
  {S.-W.}\ \bibnamefont {Kim}}, \bibinfo {author} {\bibfnamefont
  {H.}~\bibnamefont {Chang}}, \bibinfo {author} {\bibfnamefont
  {W.}~\bibnamefont {Song}}, \bibinfo {author} {\bibfnamefont {S.~S.}\
  \bibnamefont {Lee}}, \bibinfo {author} {\bibfnamefont {J.}~\bibnamefont
  {Lim}}, \bibinfo {author} {\bibfnamefont {K.-S.}\ \bibnamefont {An}}, \ and\
  \bibinfo {author} {\bibfnamefont {S.}~\bibnamefont {Myung}},\ }\href
  {\doibase https://doi.org/10.1016/j.nanoen.2019.05.017} {\bibfield  {journal}
  {\bibinfo  {journal} {Nano Energy}\ }\textbf {\bibinfo {volume} {61}},\
  \bibinfo {pages} {471 } (\bibinfo {year} {2019}{\natexlab{b}})}\BibitemShut
  {NoStop}%
\bibitem [{\citenamefont {Javvaji}\ \emph {et~al.}(2019)\citenamefont
  {Javvaji}, \citenamefont {He}, \citenamefont {Zhuang},\ and\ \citenamefont
  {Park}}]{PhysRevMaterials.3.125402}%
  \BibitemOpen
  \bibfield  {author} {\bibinfo {author} {\bibfnamefont {B.}~\bibnamefont
  {Javvaji}}, \bibinfo {author} {\bibfnamefont {B.}~\bibnamefont {He}},
  \bibinfo {author} {\bibfnamefont {X.}~\bibnamefont {Zhuang}}, \ and\ \bibinfo
  {author} {\bibfnamefont {H.~S.}\ \bibnamefont {Park}},\ }\href {\doibase
  10.1103/PhysRevMaterials.3.125402} {\bibfield  {journal} {\bibinfo  {journal}
  {Phys. Rev. Materials}\ }\textbf {\bibinfo {volume} {3}},\ \bibinfo {pages}
  {125402} (\bibinfo {year} {2019})}\BibitemShut {NoStop}%
\bibitem [{\citenamefont {Zhuang}\ \emph {et~al.}(2019)\citenamefont {Zhuang},
  \citenamefont {He}, \citenamefont {Javvaji},\ and\ \citenamefont
  {Park}}]{PhysRevB.99.054105}%
  \BibitemOpen
  \bibfield  {author} {\bibinfo {author} {\bibfnamefont {X.}~\bibnamefont
  {Zhuang}}, \bibinfo {author} {\bibfnamefont {B.}~\bibnamefont {He}}, \bibinfo
  {author} {\bibfnamefont {B.}~\bibnamefont {Javvaji}}, \ and\ \bibinfo
  {author} {\bibfnamefont {H.~S.}\ \bibnamefont {Park}},\ }\href {\doibase
  10.1103/PhysRevB.99.054105} {\bibfield  {journal} {\bibinfo  {journal} {Phys.
  Rev. B}\ }\textbf {\bibinfo {volume} {99}},\ \bibinfo {pages} {054105}
  (\bibinfo {year} {2019})}\BibitemShut {NoStop}%
\bibitem [{\citenamefont {Tenne}\ \emph {et~al.}(1992)\citenamefont {Tenne},
  \citenamefont {Margulis}, \citenamefont {Genut},\ and\ \citenamefont
  {Hodes}}]{tenne1992polyhedral}%
  \BibitemOpen
  \bibfield  {author} {\bibinfo {author} {\bibfnamefont {R.}~\bibnamefont
  {Tenne}}, \bibinfo {author} {\bibfnamefont {L.}~\bibnamefont {Margulis}},
  \bibinfo {author} {\bibfnamefont {M.~e.}\ \bibnamefont {Genut}}, \ and\
  \bibinfo {author} {\bibfnamefont {G.}~\bibnamefont {Hodes}},\ }\href@noop {}
  {\bibfield  {journal} {\bibinfo  {journal} {Nature}\ }\textbf {\bibinfo
  {volume} {360}},\ \bibinfo {pages} {444} (\bibinfo {year}
  {1992})}\BibitemShut {NoStop}%
\bibitem [{\citenamefont {Margulis}\ \emph {et~al.}(1993)\citenamefont
  {Margulis}, \citenamefont {Salitra}, \citenamefont {Tenne},\ and\
  \citenamefont {Talianker}}]{margulis1993nested}%
  \BibitemOpen
  \bibfield  {author} {\bibinfo {author} {\bibfnamefont {L.}~\bibnamefont
  {Margulis}}, \bibinfo {author} {\bibfnamefont {G.}~\bibnamefont {Salitra}},
  \bibinfo {author} {\bibfnamefont {R.}~\bibnamefont {Tenne}}, \ and\ \bibinfo
  {author} {\bibfnamefont {M.}~\bibnamefont {Talianker}},\ }\href@noop {}
  {\bibfield  {journal} {\bibinfo  {journal} {Nature}\ }\textbf {\bibinfo
  {volume} {365}},\ \bibinfo {pages} {113} (\bibinfo {year}
  {1993})}\BibitemShut {NoStop}%
\bibitem [{\citenamefont {Seifert}\ \emph {et~al.}(2000)\citenamefont
  {Seifert}, \citenamefont {Terrones}, \citenamefont {Terrones}, \citenamefont
  {Jungnickel},\ and\ \citenamefont {Frauenheim}}]{seifert2000structure}%
  \BibitemOpen
  \bibfield  {author} {\bibinfo {author} {\bibfnamefont {G.}~\bibnamefont
  {Seifert}}, \bibinfo {author} {\bibfnamefont {H.}~\bibnamefont {Terrones}},
  \bibinfo {author} {\bibfnamefont {M.}~\bibnamefont {Terrones}}, \bibinfo
  {author} {\bibfnamefont {G.}~\bibnamefont {Jungnickel}}, \ and\ \bibinfo
  {author} {\bibfnamefont {T.}~\bibnamefont {Frauenheim}},\ }\href@noop {}
  {\bibfield  {journal} {\bibinfo  {journal} {Physical Review Letters}\
  }\textbf {\bibinfo {volume} {85}},\ \bibinfo {pages} {146} (\bibinfo {year}
  {2000})}\BibitemShut {NoStop}%
\bibitem [{\citenamefont {Milo{\v{s}}evi{\'c}}\ \emph
  {et~al.}(2007)\citenamefont {Milo{\v{s}}evi{\'c}}, \citenamefont
  {Nikoli{\'c}}, \citenamefont {Dobard{\v{z}}i{\'c}}, \citenamefont
  {Damnjanovi{\'c}}, \citenamefont {Popov},\ and\ \citenamefont
  {Seifert}}]{milovsevic2007electronic}%
  \BibitemOpen
  \bibfield  {author} {\bibinfo {author} {\bibfnamefont {I.}~\bibnamefont
  {Milo{\v{s}}evi{\'c}}}, \bibinfo {author} {\bibfnamefont {B.}~\bibnamefont
  {Nikoli{\'c}}}, \bibinfo {author} {\bibfnamefont {E.}~\bibnamefont
  {Dobard{\v{z}}i{\'c}}}, \bibinfo {author} {\bibfnamefont {M.}~\bibnamefont
  {Damnjanovi{\'c}}}, \bibinfo {author} {\bibfnamefont {I.}~\bibnamefont
  {Popov}}, \ and\ \bibinfo {author} {\bibfnamefont {G.}~\bibnamefont
  {Seifert}},\ }\href@noop {} {\bibfield  {journal} {\bibinfo  {journal}
  {Physical Review B}\ }\textbf {\bibinfo {volume} {76}},\ \bibinfo {pages}
  {233414} (\bibinfo {year} {2007})}\BibitemShut {NoStop}%
\bibitem [{\citenamefont {Tenne}\ and\ \citenamefont
  {Redlich}(2010)}]{tenne2010recent}%
  \BibitemOpen
  \bibfield  {author} {\bibinfo {author} {\bibfnamefont {R.}~\bibnamefont
  {Tenne}}\ and\ \bibinfo {author} {\bibfnamefont {M.}~\bibnamefont
  {Redlich}},\ }\href@noop {} {\bibfield  {journal} {\bibinfo  {journal}
  {Chemical Society Reviews}\ }\textbf {\bibinfo {volume} {39}},\ \bibinfo
  {pages} {1423} (\bibinfo {year} {2010})}\BibitemShut {NoStop}%
\bibitem [{\citenamefont {Zibouche}\ \emph {et~al.}(2012)\citenamefont
  {Zibouche}, \citenamefont {Kuc},\ and\ \citenamefont
  {Heine}}]{zibouche2012layers}%
  \BibitemOpen
  \bibfield  {author} {\bibinfo {author} {\bibfnamefont {N.}~\bibnamefont
  {Zibouche}}, \bibinfo {author} {\bibfnamefont {A.}~\bibnamefont {Kuc}}, \
  and\ \bibinfo {author} {\bibfnamefont {T.}~\bibnamefont {Heine}},\
  }\href@noop {} {\bibfield  {journal} {\bibinfo  {journal} {The European
  Physical Journal B}\ }\textbf {\bibinfo {volume} {85}},\ \bibinfo {pages}
  {49} (\bibinfo {year} {2012})}\BibitemShut {NoStop}%
\bibitem [{\citenamefont {Artyukhov}\ \emph {et~al.}(0)\citenamefont
  {Artyukhov}, \citenamefont {Gupta}, \citenamefont {Kutana},\ and\
  \citenamefont {Yakobson}}]{doi:10.1021/acs.nanolett.9b05345}%
  \BibitemOpen
  \bibfield  {author} {\bibinfo {author} {\bibfnamefont {V.~I.}\ \bibnamefont
  {Artyukhov}}, \bibinfo {author} {\bibfnamefont {S.}~\bibnamefont {Gupta}},
  \bibinfo {author} {\bibfnamefont {A.}~\bibnamefont {Kutana}}, \ and\ \bibinfo
  {author} {\bibfnamefont {B.~I.}\ \bibnamefont {Yakobson}},\ }\href {\doibase
  10.1021/acs.nanolett.9b05345} {\bibfield  {journal} {\bibinfo  {journal}
  {Nano Letters}\ }\textbf {\bibinfo {volume} {0}},\ \bibinfo {pages} {null}
  (\bibinfo {year} {0})},\ \bibinfo {note} {pMID: 32155086}\BibitemShut
  {NoStop}%
\bibitem [{\citenamefont {Junquera}\ \emph {et~al.}(2003)\citenamefont
  {Junquera}, \citenamefont {Zimmer}, \citenamefont {Ordej{\'o}n},\ and\
  \citenamefont {Ghosez}}]{junquera2003first}%
  \BibitemOpen
  \bibfield  {author} {\bibinfo {author} {\bibfnamefont {J.}~\bibnamefont
  {Junquera}}, \bibinfo {author} {\bibfnamefont {M.}~\bibnamefont {Zimmer}},
  \bibinfo {author} {\bibfnamefont {P.}~\bibnamefont {Ordej{\'o}n}}, \ and\
  \bibinfo {author} {\bibfnamefont {P.}~\bibnamefont {Ghosez}},\ }\href@noop {}
  {\bibfield  {journal} {\bibinfo  {journal} {Physical Review B}\ }\textbf
  {\bibinfo {volume} {67}},\ \bibinfo {pages} {155327} (\bibinfo {year}
  {2003})}\BibitemShut {NoStop}%
\bibitem [{\citenamefont {Junquera}\ \emph {et~al.}(2007)\citenamefont
  {Junquera}, \citenamefont {Cohen},\ and\ \citenamefont {Rabe}}]{nanosmooth}%
  \BibitemOpen
  \bibfield  {author} {\bibinfo {author} {\bibfnamefont {J.}~\bibnamefont
  {Junquera}}, \bibinfo {author} {\bibfnamefont {M.~H.}\ \bibnamefont {Cohen}},
  \ and\ \bibinfo {author} {\bibfnamefont {K.~M.}\ \bibnamefont {Rabe}},\
  }\href@noop {} {\bibfield  {journal} {\bibinfo  {journal} {Journal of
  Physics: Condensed Matter}\ }\textbf {\bibinfo {volume} {19}},\ \bibinfo
  {pages} {213203} (\bibinfo {year} {2007})}\BibitemShut {NoStop}%
\bibitem [{\citenamefont {Stengel}\ \emph {et~al.}(2011)\citenamefont
  {Stengel}, \citenamefont {Aguado-Puente}, \citenamefont {Spaldin},\ and\
  \citenamefont {Junquera}}]{stengel2011band}%
  \BibitemOpen
  \bibfield  {author} {\bibinfo {author} {\bibfnamefont {M.}~\bibnamefont
  {Stengel}}, \bibinfo {author} {\bibfnamefont {P.}~\bibnamefont
  {Aguado-Puente}}, \bibinfo {author} {\bibfnamefont {N.~A.}\ \bibnamefont
  {Spaldin}}, \ and\ \bibinfo {author} {\bibfnamefont {J.}~\bibnamefont
  {Junquera}},\ }\href@noop {} {\bibfield  {journal} {\bibinfo  {journal}
  {Physical Review B}\ }\textbf {\bibinfo {volume} {83}},\ \bibinfo {pages}
  {235112} (\bibinfo {year} {2011})}\BibitemShut {NoStop}%
\bibitem [{\citenamefont {Soler}\ \emph {et~al.}(2002)\citenamefont {Soler},
  \citenamefont {Artacho}, \citenamefont {Gale}, \citenamefont {Garc{\'\i}a},
  \citenamefont {Junquera}, \citenamefont {Ordej{\'o}n},\ and\ \citenamefont
  {S{\'a}nchez-Portal}}]{siesta}%
  \BibitemOpen
  \bibfield  {author} {\bibinfo {author} {\bibfnamefont {J.~M.}\ \bibnamefont
  {Soler}}, \bibinfo {author} {\bibfnamefont {E.}~\bibnamefont {Artacho}},
  \bibinfo {author} {\bibfnamefont {J.~D.}\ \bibnamefont {Gale}}, \bibinfo
  {author} {\bibfnamefont {A.}~\bibnamefont {Garc{\'\i}a}}, \bibinfo {author}
  {\bibfnamefont {J.}~\bibnamefont {Junquera}}, \bibinfo {author}
  {\bibfnamefont {P.}~\bibnamefont {Ordej{\'o}n}}, \ and\ \bibinfo {author}
  {\bibfnamefont {D.}~\bibnamefont {S{\'a}nchez-Portal}},\ }\href@noop {}
  {\bibfield  {journal} {\bibinfo  {journal} {Journal of Physics: Condensed
  Matter}\ }\textbf {\bibinfo {volume} {14}},\ \bibinfo {pages} {2745}
  (\bibinfo {year} {2002})}\BibitemShut {NoStop}%
\bibitem [{\citenamefont {Garc{\'\i}a}\ \emph {et~al.}(2018)\citenamefont
  {Garc{\'\i}a}, \citenamefont {Verstraete}, \citenamefont {Pouillon},\ and\
  \citenamefont {Junquera}}]{psml}%
  \BibitemOpen
  \bibfield  {author} {\bibinfo {author} {\bibfnamefont {A.}~\bibnamefont
  {Garc{\'\i}a}}, \bibinfo {author} {\bibfnamefont {M.~J.}\ \bibnamefont
  {Verstraete}}, \bibinfo {author} {\bibfnamefont {Y.}~\bibnamefont
  {Pouillon}}, \ and\ \bibinfo {author} {\bibfnamefont {J.}~\bibnamefont
  {Junquera}},\ }\href@noop {} {\bibfield  {journal} {\bibinfo  {journal}
  {Computer Physics Communications}\ }\textbf {\bibinfo {volume} {227}},\
  \bibinfo {pages} {51} (\bibinfo {year} {2018})}\BibitemShut {NoStop}%
\bibitem [{\citenamefont {Junquera}()}]{javier_psml}%
  \BibitemOpen
  \bibfield  {author} {\bibinfo {author} {\bibfnamefont {J.}~\bibnamefont
  {Junquera}},\ }\href@noop {} {\enquote {\bibinfo {title} {{How to generate
  PSML pseudopotentials and run SIESTA and ABINIT with the same pseudo}},}\
  }\bibinfo {howpublished}
  {\url{https://personales.unican.es/junqueraj/JavierJunquera_files/Metodos/Pseudos/Siesta-abinit/Siesta-Abinit-PSML.pdf}}\BibitemShut
  {NoStop}%
\bibitem [{\citenamefont {Hamann}(2013)}]{norm_conserving}%
  \BibitemOpen
  \bibfield  {author} {\bibinfo {author} {\bibfnamefont {D.}~\bibnamefont
  {Hamann}},\ }\href@noop {} {\bibfield  {journal} {\bibinfo  {journal}
  {Physical Review B}\ }\textbf {\bibinfo {volume} {88}},\ \bibinfo {pages}
  {085117} (\bibinfo {year} {2013})}\BibitemShut {NoStop}%
\bibitem [{\citenamefont {Van~Setten}\ \emph {et~al.}(2018)\citenamefont
  {Van~Setten}, \citenamefont {Giantomassi}, \citenamefont {Bousquet},
  \citenamefont {Verstraete}, \citenamefont {Hamann}, \citenamefont {Gonze},\
  and\ \citenamefont {Rignanese}}]{pseudodojo}%
  \BibitemOpen
  \bibfield  {author} {\bibinfo {author} {\bibfnamefont {M.}~\bibnamefont
  {Van~Setten}}, \bibinfo {author} {\bibfnamefont {M.}~\bibnamefont
  {Giantomassi}}, \bibinfo {author} {\bibfnamefont {E.}~\bibnamefont
  {Bousquet}}, \bibinfo {author} {\bibfnamefont {M.~J.}\ \bibnamefont
  {Verstraete}}, \bibinfo {author} {\bibfnamefont {D.~R.}\ \bibnamefont
  {Hamann}}, \bibinfo {author} {\bibfnamefont {X.}~\bibnamefont {Gonze}}, \
  and\ \bibinfo {author} {\bibfnamefont {G.-M.}\ \bibnamefont {Rignanese}},\
  }\href@noop {} {\bibfield  {journal} {\bibinfo  {journal} {Computer Physics
  Communications}\ }\textbf {\bibinfo {volume} {226}},\ \bibinfo {pages} {39}
  (\bibinfo {year} {2018})}\BibitemShut {NoStop}%
\bibitem [{\citenamefont {Artacho}\ \emph {et~al.}(1999)\citenamefont
  {Artacho}, \citenamefont {S{\'a}nchez-Portal}, \citenamefont {Ordej{\'o}n},
  \citenamefont {Garcia},\ and\ \citenamefont {Soler}}]{siesta_2}%
  \BibitemOpen
  \bibfield  {author} {\bibinfo {author} {\bibfnamefont {E.}~\bibnamefont
  {Artacho}}, \bibinfo {author} {\bibfnamefont {D.}~\bibnamefont
  {S{\'a}nchez-Portal}}, \bibinfo {author} {\bibfnamefont {P.}~\bibnamefont
  {Ordej{\'o}n}}, \bibinfo {author} {\bibfnamefont {A.}~\bibnamefont {Garcia}},
  \ and\ \bibinfo {author} {\bibfnamefont {J.~M.}\ \bibnamefont {Soler}},\
  }\href@noop {} {\bibfield  {journal} {\bibinfo  {journal} {physica status
  solidi (b)}\ }\textbf {\bibinfo {volume} {215}},\ \bibinfo {pages} {809}
  (\bibinfo {year} {1999})}\BibitemShut {NoStop}%
\bibitem [{\citenamefont {Corsetti}\ \emph {et~al.}(2013)\citenamefont
  {Corsetti}, \citenamefont {Fern{\'{a}}ndez-Serra}, \citenamefont {Soler},\
  and\ \citenamefont {Artacho}}]{basis_water}%
  \BibitemOpen
  \bibfield  {author} {\bibinfo {author} {\bibfnamefont {F.}~\bibnamefont
  {Corsetti}}, \bibinfo {author} {\bibfnamefont {M.-V.}\ \bibnamefont
  {Fern{\'{a}}ndez-Serra}}, \bibinfo {author} {\bibfnamefont {J.~M.}\
  \bibnamefont {Soler}}, \ and\ \bibinfo {author} {\bibfnamefont
  {E.}~\bibnamefont {Artacho}},\ }\href {\doibase
  10.1088/0953-8984/25/43/435504} {\bibfield  {journal} {\bibinfo  {journal}
  {Journal of Physics: Condensed Matter}\ }\textbf {\bibinfo {volume} {25}},\
  \bibinfo {pages} {435504} (\bibinfo {year} {2013})}\BibitemShut {NoStop}%
\bibitem [{\citenamefont {Perdew}\ and\ \citenamefont {Wang}(1992)}]{pw92}%
  \BibitemOpen
  \bibfield  {author} {\bibinfo {author} {\bibfnamefont {J.~P.}\ \bibnamefont
  {Perdew}}\ and\ \bibinfo {author} {\bibfnamefont {Y.}~\bibnamefont {Wang}},\
  }\href {\doibase 10.1103/PhysRevB.45.13244} {\bibfield  {journal} {\bibinfo
  {journal} {Phys. Rev. B}\ }\textbf {\bibinfo {volume} {45}},\ \bibinfo
  {pages} {13244} (\bibinfo {year} {1992})}\BibitemShut {NoStop}%
\bibitem [{\citenamefont {Perdew}\ \emph {et~al.}(1996)\citenamefont {Perdew},
  \citenamefont {Burke},\ and\ \citenamefont {Ernzerhof}}]{pbe}%
  \BibitemOpen
  \bibfield  {author} {\bibinfo {author} {\bibfnamefont {J.~P.}\ \bibnamefont
  {Perdew}}, \bibinfo {author} {\bibfnamefont {K.}~\bibnamefont {Burke}}, \
  and\ \bibinfo {author} {\bibfnamefont {M.}~\bibnamefont {Ernzerhof}},\ }\href
  {\doibase 10.1103/PhysRevLett.77.3865} {\bibfield  {journal} {\bibinfo
  {journal} {Phys. Rev. Lett.}\ }\textbf {\bibinfo {volume} {77}},\ \bibinfo
  {pages} {3865} (\bibinfo {year} {1996})}\BibitemShut {NoStop}%
\bibitem [{\citenamefont {Monkhorst}\ and\ \citenamefont {Pack}(1976)}]{mp}%
  \BibitemOpen
  \bibfield  {author} {\bibinfo {author} {\bibfnamefont {H.~J.}\ \bibnamefont
  {Monkhorst}}\ and\ \bibinfo {author} {\bibfnamefont {J.~D.}\ \bibnamefont
  {Pack}},\ }\href@noop {} {\bibfield  {journal} {\bibinfo  {journal} {Physical
  Review B}\ }\textbf {\bibinfo {volume} {13}},\ \bibinfo {pages} {5188}
  (\bibinfo {year} {1976})}\BibitemShut {NoStop}%
\bibitem [{\citenamefont {Neugebauer}\ and\ \citenamefont
  {Scheffler}(1992)}]{dipole_correction_1}%
  \BibitemOpen
  \bibfield  {author} {\bibinfo {author} {\bibfnamefont {J.}~\bibnamefont
  {Neugebauer}}\ and\ \bibinfo {author} {\bibfnamefont {M.}~\bibnamefont
  {Scheffler}},\ }\href {\doibase 10.1103/PhysRevB.46.16067} {\bibfield
  {journal} {\bibinfo  {journal} {Phys. Rev. B}\ }\textbf {\bibinfo {volume}
  {46}},\ \bibinfo {pages} {16067} (\bibinfo {year} {1992})}\BibitemShut
  {NoStop}%
\bibitem [{\citenamefont {Bengtsson}(1999)}]{dipole_correction_2}%
  \BibitemOpen
  \bibfield  {author} {\bibinfo {author} {\bibfnamefont {L.}~\bibnamefont
  {Bengtsson}},\ }\href {\doibase 10.1103/PhysRevB.59.12301} {\bibfield
  {journal} {\bibinfo  {journal} {Phys. Rev. B}\ }\textbf {\bibinfo {volume}
  {59}},\ \bibinfo {pages} {12301} (\bibinfo {year} {1999})}\BibitemShut
  {NoStop}%
\bibitem [{\citenamefont {Kantorovich}(1999)}]{dipole_correction_3}%
  \BibitemOpen
  \bibfield  {author} {\bibinfo {author} {\bibfnamefont {L.~N.}\ \bibnamefont
  {Kantorovich}},\ }\href {\doibase 10.1103/PhysRevB.60.15476} {\bibfield
  {journal} {\bibinfo  {journal} {Phys. Rev. B}\ }\textbf {\bibinfo {volume}
  {60}},\ \bibinfo {pages} {15476} (\bibinfo {year} {1999})}\BibitemShut
  {NoStop}%
\bibitem [{\citenamefont {Meyer}\ and\ \citenamefont
  {Vanderbilt}(2001)}]{dipole_correction_4}%
  \BibitemOpen
  \bibfield  {author} {\bibinfo {author} {\bibfnamefont {B.}~\bibnamefont
  {Meyer}}\ and\ \bibinfo {author} {\bibfnamefont {D.}~\bibnamefont
  {Vanderbilt}},\ }\href@noop {} {\bibfield  {journal} {\bibinfo  {journal}
  {Physical Review B}\ }\textbf {\bibinfo {volume} {63}},\ \bibinfo {pages}
  {205426} (\bibinfo {year} {2001})}\BibitemShut {NoStop}%
\bibitem [{Note1()}]{Note1}%
  \BibitemOpen
  \bibinfo {note} {Although this is not necessary if the monolayers are
  correctly relaxed as they should be non-polar.}\BibitemShut {Stop}%
\bibitem [{\citenamefont {Rasmussen}\ and\ \citenamefont
  {Thygesen}(2015)}]{computational_database_2d}%
  \BibitemOpen
  \bibfield  {author} {\bibinfo {author} {\bibfnamefont {F.~A.}\ \bibnamefont
  {Rasmussen}}\ and\ \bibinfo {author} {\bibfnamefont {K.~S.}\ \bibnamefont
  {Thygesen}},\ }\href@noop {} {\bibfield  {journal} {\bibinfo  {journal} {The
  Journal of Physical Chemistry C}\ }\textbf {\bibinfo {volume} {119}},\
  \bibinfo {pages} {13169} (\bibinfo {year} {2015})}\BibitemShut {NoStop}%
\bibitem [{\citenamefont {Rutter}(2018)}]{c2x}%
  \BibitemOpen
  \bibfield  {author} {\bibinfo {author} {\bibfnamefont {M.~J.}\ \bibnamefont
  {Rutter}},\ }\href@noop {} {\bibfield  {journal} {\bibinfo  {journal}
  {Computer Physics Communications}\ }\textbf {\bibinfo {volume} {225}},\
  \bibinfo {pages} {174} (\bibinfo {year} {2018})}\BibitemShut {NoStop}%
\bibitem [{\citenamefont {Edgcombe}\ \emph {et~al.}(2019)\citenamefont
  {Edgcombe}, \citenamefont {Masur}, \citenamefont {Linscott}, \citenamefont
  {Whaley-Baldwin},\ and\ \citenamefont {Barnes}}]{jwb}%
  \BibitemOpen
  \bibfield  {author} {\bibinfo {author} {\bibfnamefont {C.}~\bibnamefont
  {Edgcombe}}, \bibinfo {author} {\bibfnamefont {S.}~\bibnamefont {Masur}},
  \bibinfo {author} {\bibfnamefont {E.}~\bibnamefont {Linscott}}, \bibinfo
  {author} {\bibfnamefont {J.}~\bibnamefont {Whaley-Baldwin}}, \ and\ \bibinfo
  {author} {\bibfnamefont {C.}~\bibnamefont {Barnes}},\ }\href@noop {}
  {\bibfield  {journal} {\bibinfo  {journal} {Ultramicroscopy}\ }\textbf
  {\bibinfo {volume} {198}},\ \bibinfo {pages} {26} (\bibinfo {year}
  {2019})}\BibitemShut {NoStop}%
\bibitem [{\citenamefont {Moghadas}\ \emph {et~al.}(2019)\citenamefont
  {Moghadas}, \citenamefont {Poursamad}, \citenamefont {Sahrai},\ and\
  \citenamefont {Emdadi}}]{moghadas2019flexoelectric}%
  \BibitemOpen
  \bibfield  {author} {\bibinfo {author} {\bibfnamefont {F.}~\bibnamefont
  {Moghadas}}, \bibinfo {author} {\bibfnamefont {J.}~\bibnamefont {Poursamad}},
  \bibinfo {author} {\bibfnamefont {M.}~\bibnamefont {Sahrai}}, \ and\ \bibinfo
  {author} {\bibfnamefont {M.}~\bibnamefont {Emdadi}},\ }\href@noop {}
  {\bibfield  {journal} {\bibinfo  {journal} {The European Physical Journal E}\
  }\textbf {\bibinfo {volume} {42}},\ \bibinfo {pages} {103} (\bibinfo {year}
  {2019})}\BibitemShut {NoStop}%
\bibitem [{\citenamefont {Lu}\ \emph {et~al.}(2004)\citenamefont {Lu},
  \citenamefont {Li}, \citenamefont {Rotkin}, \citenamefont {Ravaioli},\ and\
  \citenamefont {Schulten}}]{lu2004finite}%
  \BibitemOpen
  \bibfield  {author} {\bibinfo {author} {\bibfnamefont {D.}~\bibnamefont
  {Lu}}, \bibinfo {author} {\bibfnamefont {Y.}~\bibnamefont {Li}}, \bibinfo
  {author} {\bibfnamefont {S.~V.}\ \bibnamefont {Rotkin}}, \bibinfo {author}
  {\bibfnamefont {U.}~\bibnamefont {Ravaioli}}, \ and\ \bibinfo {author}
  {\bibfnamefont {K.}~\bibnamefont {Schulten}},\ }\href@noop {} {\bibfield
  {journal} {\bibinfo  {journal} {Nano Letters}\ }\textbf {\bibinfo {volume}
  {4}},\ \bibinfo {pages} {2383} (\bibinfo {year} {2004})}\BibitemShut
  {NoStop}%
\bibitem [{\citenamefont {Li}\ \emph {et~al.}(2005)\citenamefont {Li},
  \citenamefont {Lu}, \citenamefont {Schulten}, \citenamefont {Ravaioli},\ and\
  \citenamefont {Rotkin}}]{li2005screening}%
  \BibitemOpen
  \bibfield  {author} {\bibinfo {author} {\bibfnamefont {Y.}~\bibnamefont
  {Li}}, \bibinfo {author} {\bibfnamefont {D.}~\bibnamefont {Lu}}, \bibinfo
  {author} {\bibfnamefont {K.}~\bibnamefont {Schulten}}, \bibinfo {author}
  {\bibfnamefont {U.}~\bibnamefont {Ravaioli}}, \ and\ \bibinfo {author}
  {\bibfnamefont {S.~V.}\ \bibnamefont {Rotkin}},\ }\href@noop {} {\bibfield
  {journal} {\bibinfo  {journal} {Journal of computational electronics}\
  }\textbf {\bibinfo {volume} {4}},\ \bibinfo {pages} {161} (\bibinfo {year}
  {2005})}\BibitemShut {NoStop}%
\end{thebibliography}

%

\end{document}